**Effects of rearing density on growth, digestive conditions, welfare indicators and gut bacterial community of gilthead sea bream (*Sparus aurata*, L. 1758) fed different fishmeal and fish oil dietary levels**

Luca Parma[a]*, Nicole Francesca Pelusio[a], Enric Gisbert[b], Maria Angeles Esteban[c], Federica D'Amico[d], Matteo Soverini[d], Marco Candela[d], Francesco Dondi[a], Pier Paolo Gatta[a], Alessio Bonaldo[a]

[a]Department of Veterinary Medical Sciences, University of Bologna, Via Tolara di Sopra 50, 40064 Ozzano Emilia, Italy

[b]IRTA – Sant Carles de la Ràpita, Programa d'Aqüicultura, Crta. del Poble Nou km 5.5, 43540 Sant Carles de la Ràpita, Spain

[c]Department of Cell Biology and Histology, Faculty of Biology, Campus Regional de Excelencia Internacional "Campus Mare Nostrum", University of Murcia, 30100 Murcia, Spain

[d]Unit of Microbial Ecology of Health, Department of Pharmacy and Biotechnology, University of Bologna, Via Belmeloro 6, 40126 Bologna, Italy

*Corresponding author*: Luca Parma, Department of Veterinary Medical Sciences, University of Bologna, Viale Vespucci 2, 47042 Cesenatico, FC, Italy. *Tel*.: +39 0547 338931; *Fax*: +39 0547 338941

E-mail address: luca.parma@unibo.it (L. Parma)

**Abstract**

In Mediterranean aquaculture, significant advances have been made towards a reduction of marine-derived ingredients in aquafeed formulation, as well as in defining the effect on how environmental factors such as rearing density interact with fish health. Little research, however, has examined the interaction between rearing density and dietary composition on main key performance indicators, physiological processes and gut bacterial community. A study was undertaken, therefore to assess growth response, digestive enzyme activity, humoral immunity on skin mucus, plasma biochemistry and gut microbiota of gilthead sea bream (*Sparus aurata*, L. 1758) reared at high (HD, 36-44 kg m$^{-3}$) and low (LD, 12-15 kg m$^{-3}$) final stocking densities and fed high (FM30/FO15, 30% fishmeal FM, 15% fish oil, FO) and low (FM10/FO3; 10% FM and 3% FO) FM and FO levels. Isonitrogenous and isolipidic extruded diets were fed to triplicate fish groups (initial weight: 96.2 g) to overfeeding over 98 days. The densities tested had no major effects on overall growth and feed efficiency of sea bream reared at high or low FM and FO dietary level. However, HD seems to reduce feed intake compared to LD mainly in fish fed FM30/FO15. Results of digestive enzyme activity indicated a comparable digestive efficiency among rearing densities and within each dietary treatment even if intestinal brush border enzymes appeared to be more influenced by stocking density compared to gastric and pancreatic enzymes. Plasma parameters related to nutritional and physiological conditions were not affected by rearing densities under both nutritional conditions a similar observation was also achieved through the study of lysozyme, protease, antiprotease and total protein determination in skin mucus, however; in this case lysozyme was slightly reduced at HD. For the first time on this species, the effect of rearing density on gut bacterial community was studied. Different response in relation to dietary treatment under HD and LD were detected. Low FM-FO diet maintained steady the biodiversity of the gut bacterial community between LD and HD conditions while fish fed high FM-FO level showed a reduced biodiversity at HD. According to the results, it seems feasible to rear gilthead sea bream at the on-growing phase at a density up to 36-44 kg m$^{-3}$ with low or high FM-FO diet without negatively affecting growth, feed efficiency, welfare condition and gut bacterial community.

**Introduction**

Despite the considerable advances addressing the study of nutritional requirements and sustainable feed ingredients in fish, which have resulted in a deep knowledge about the optimal composition of aquafeeds for Mediterranean fish species, technical performance indicators such as growth, feed utilization and survival in Mediterranean aquaculture have not improved over the last decade. The intensification of production systems and their possible effects on stress and welfare or the less explored interaction between nutrition, feeding management and suboptimal environmental conditions may have contributed to this stagnation. Among stress factors, inadequate rearing density has been recognized as a source of chronic stress in fish species which could affect physiological processes such as osmoregulation or immune competence, mobilization of energy sources and alterations in behaviour, which are generally translated into a decreased feed intake, reduced feed efficiency and decreased growth performance (Ellis et al., 2002; Tort et al., 2011). In gilthead sea bream (*Sparus aurata*), several studies have evaluated the effects of stocking density on growth and fish health. In juveniles, Canario et al. (1998) found that growth was negatively correlated to stocking density when fish were reared at a final stocking density of 16.8 kg m$^{-3}$ compared to 2.4 kg m$^{-3}$, while Montero et al. (1999) did not find an effect on growth and feed intake when specimens (22-85 g) were reared up to 40.8 kg m$^{-3}$, even if a negative effect on plasma and serum parameters were detected. More recently high stocking density (final density 57 kg m$^{-3}$) decreased growth performance, feed intake and feed efficiency of gilthead sea bream (12-58 g) in comparison to lower density 5-26 kg m$^{-3}$ (Diogenes et al., 2019). In addition, in adult fish (272-425g) rearing density was increased up to 20 kg m$^{-3}$ without affecting physiological parameters and growth, when oxygen level was maintained above 70% of the saturation level (Araujo-Luna et al., 2018). Concerning the effect of rearing density on welfare in this species, several studies have elucidated the effect on different physiological parameters, including plasma parameters, neuroendocrine factors, skin mucus biomarkers, liver proteome, carbohydrate metabolism of several tissues and behavioural studies (Montero et al., 1999; Sangiao-Alvarellos et al., 2005; Mancera et al., 2008; Alves et al., 2010; Sánchez-Muros et al., 2017; Guardiola et al., 2018; Skrzynska et al., 2018; Diógenes et al., 2019). Most of those studies were conducted using standard diets and whether these density-associated changes in performance and welfare are consistent when fish are fed current low fishmeal (FM) and fish oil (FO) diets remains little investigated (Wong et al., 2013). In addition, only a few studies in fish species have evaluated whether the interaction between stocking density and diet composition may affect gut microbiota and none of these have been evaluated in gilthead sea bream. The exposure to stress factors can impact the gut microbiome community profile by altering the relative proportions of the main microbiota phyla (Galley et al., 2014), while a recent study on blunt snout bream (*Megalobrama amblycephala*) provided new evidence that the gut microbiome might be involved in the response to crowding and consequently to the adaptation of fish to environmental stressors (Du et al., 2019). The aim of the present study was to explore the effect of high and low rearing density on growth, digestive enzyme activity, plasma biochemistry, humoral immunity of skin mucus and gut microbiome structure during the on-growing of gilthead sea bream fed low and high FM and FO dietary levels.

**Materials and methods**

*2.1 Experimental diets*

Ingredients and proximate composition of the experimental diets are presented in Table 1. Two isonitrogenous (46% protein) and isolipidic (17% lipid) diets were formulated to contain high and low FM and FO dietary levels (FM30/FO15 and FM10/FO3; 30% FM, 15% FO and 10% FM and 3% FO, respectively). Diets were formulated with FM and with a mixture of vegetable ingredients currently used for sea bream in aquafeed (Parma et al., 2016). The diets were produced via extrusion (pellet size = 4.0 mm) by SPAROS Lda (Portugal).

*2.2 Fish density and rearing*

The experiment was carried out at the Laboratory of Aquaculture, Department of Veterinary Medical Sciences of the University of Bologna (Cesenatico, Italy). Gilthead sea bream were obtained from the fish farm Cosa s.r.l (Orbello, GR) and adapted to the laboratory facilities for 10 days before the beginning of the trial. Afterwards, two rearing densities (low density and high density, LD and HD, respectively) were established by randomly distributing 40 and 120 fish per tank (96.2 ± 2.1g) in six 800L tanks corresponding to an initial density of 4.8 and 14.4 kg m$^{-3}$, respectively (Table 2).

Each diet was administered to triplicate tanks at both rearing densities over 98 days. Tanks were provided with natural seawater and connected to a closed recirculation system (overall water volume: 15 m$^{-3}$). The rearing system consisted of a mechanical sand filter (PTK 1200, Astralpool, Barcelona, Spain), ultraviolet lights (PE 25mJ cm$^{-2}$: 32 m$^{-3}$ h$^{-1}$, Blaufish, Barcelona, Spain) and a biofilter (PTK 1200, Astralpool, Barcelona, Spain). The water exchange rate within each tank was 100% every hour, while the overall water renewal amount in the system was 5% daily. During the trial, the temperature was kept at 24 ± 1.0 °C and the photoperiod was maintained at 12 h light and 12 h dark by means of artificial light. The oxygen level was kept constant (8.0 ± 1.0 mg L$^{-1}$) through a liquid oxygen system regulated by a software programme (B&G Sinergia snc, Chioggia, Italy). Ammonia (total ammonia nitrogen ≤ 0.1 mg L$^{-1}$) and nitrite (≤ 0.2 mg L$^{-1}$) were daily monitored spectrophotometrically (Spectroquant Nova 60, Merck, Lab business, Darmstadt, Germany) while salinity (30 g L$^{-1}$) was measured by a salt refractometer (106 ATC). Sodium bicarbonate was added on a daily basis to keep pH constant at 7.8–8.0. Fish were fed *ad libitum* twice a day (8:30, 16:30) for six days a week (one meal on Sundays) via automatic feeders using an overfeeding approach with a daily feeding ration 10% higher than the daily ingested ration of the previous days as reported by Bonvini et al. (2018a). Each meal lasted 1 h, after which the uneaten pellets of each tank were collected, dried overnight at 105°C, and weighted for overall calculation.

*2.3 Sampling*

At the beginning and at the end of the experiment, all the fish in each tank were anaesthetised by 2-phenoxyethanol at 300 mg $L^{-1}$ and individually weighed. The proximate composition of the carcasses was determined at the beginning of the trial on a pooled sample of 10 fish and on a pooled sample of 5 fish per tank at the end of the trial.

At the end of the trial, for the assessment of the specific activity of gastric (pepsin) and pancreatic (trypsin, chymotrypsin, total alkaline proteases, α-amylase and bile salt-activated lipase) digestive enzymes, 3 fish per tank (n = 9 fish per diet treatment) at 5 hours post meal (hpm) were randomly sampled, euthanized with overdose anaesthetic and immediately eviscerated. The alimentary tract was dissected, adherent adipose and connective tissues carefully removed and the gastrointestinal tract was stored at −80 °C until their analysis. For the analysis of intestinal enzymes (alkaline phosphatase, maltase, aminopeptidase-N and leucine-alanine peptidase), 3 fish per tank were sampled at 8 hpm, at the same time, after fish dissection, anterior and posterior intestines were dissected and stored at −80 °C until their analysis. Sampling times were selected in order to maximize pancreatic enzyme levels in the stomach and anterior region of the intestine coinciding with their maximal secretion into the gut from the exocrine pancreas due to the presence of feed in the gut, while the activity of intestinal enzymes was measured at the end of the digestion process (Deguara et al., 2013).The measurements of digestive enzymes was then obtained by pooling the 3 fish sampled per tank during the analyses, as the tank was considered as the experimental unit and not the organism. At the same time, digesta content from posterior intestine (n = 15 fish per diet treatment, n = 5 fish per replicate) was also individually sampled and immediately stored at −80 °C for gut microbiota analysis according to Parma et al. (2016).

For the assessment of plasma biochemistry, blood from 5 fish per tank (n=15 fish per diet treatment) was collected from the caudal vein. Samples were then centrifuged (3000 x *g*, 10 min, 4°C) and plasma aliquots were stored at −80 °C until analysis (Bonvini et al., 2018b). Skin mucus samples were collected from 8 fish per tank according to the method of Guardiola et al. (2014). Briefly, skin mucus was collected by gently scraping the dorsolateral surface of specimens using a cell scraper, taking care to avoid contamination with urino-genital and intestinal excretions. Collected mucus samples were then stored at −80 °C until analyses.

All experimental procedures were evaluated and approved by the Ethical-Scientific Committee for Animal Experimentation of the University of Bologna, in accordance with European directive 2010/63/UE on the protection of animals used for scientific purposes.

*2.4 Calculations*

The following formulae were used to calculate different performance parameters: specific growth rate (SGR) (% day$^{-1}$) = 100 * (ln FBW- ln IBW) / days (where FBW and IBW represent the final and the initial body weights, respectively). Feed Intake (FI) (g kg ABW$^{-1}$ day$^{-1}$)=((1000 ∗ total ingestion)/(ABW))/days)) (where average body weight, ABW=(IBW+FBW)/2. Feed conversion ratio (FCR) = feed intake / weight gain. Protein efficiency rate (PER) = (FBW − IBW) / protein intake. Gross protein efficiency (GPE) (%) = 100 * [(% final body protein * FBW) - (% initial body protein * IBW)] / total protein intake fish. Gross lipid efficiency (GLE) = 100 * [(final body lipid (%) * FBW) - (initial body lipid (%) *IBW)] / total lipid intake fish. Lipid efficiency ratio (LER) = [(FBW-IBW)/lipid intake].

*2.5 Proximate composition analysis*

Diets and whole body of sampled fish were analysed for proximate composition. Moisture content was obtained by weight loss after drying samples in a stove at 105 °C until a constant weight was achieved. Crude protein was determined as total nitrogen (N) by using the Kjeldahl method and multiplying N by 6.25. Total lipids were determined according to Bligh and Dyer's (1959) extraction method. Ash content was estimated by incineration to a constant weight in a muffle oven at 450 °C. Gross energy was determined by a calorimetric bomb (Adiabatic Calorimetric Bomb Parr 1261; PARR Instrument, IL, U.S.A).

*2.6 Digestive enzyme activity*

Determination of pancreatic (α-amylase, bile salt-activated lipase, total alkaline proteases), gastric (pepsin) and intestinal (alkaline phosphatase, aminopeptidase-N, maltase and leucine-alanine peptidase) digestive enzymes were based on methods previously described by Gisbert et al. (2009). In addition, spectrophotometric analyses were performed as recommended by Solovyev and Gisbert (2016) in order to prevent sample deterioration. In brief, the stomach and pyloric caeca samples (including 1 cm of anterior intestine) were homogenized in 5 volumes (ww/v) of distilled water at 4 °C for 1 min followed by a sonication process of 30 sec. After a centrifugation (9,000 x *g* for 10 min at 4 °C), the supernatant was collected, aliquoted and stored at −20°C for the quantification of gastric and pancreatic digestive enzymes.

Regarding intestinal enzymes, the anterior and posterior intestine samples were homogenized in 30 volumes (w/v) of ice-cold Mannitol (50 mM), Tris-HCl buffer (2 mM) pH 7.0, at a maximum speed for 30 s (IKA, Ultra-turrax®, USA), then 100 μL of 0.1M $CaCl_2$ was added to the homogenate, stirred and centrifuged (9,000 x *g* for 10 min at 4 °C). A fraction of the supernatant was collected and stored at −20 °C for the leucine-alanine peptidase (LAP) activity quantification. After a second centrifugation (3,400 x *g* for 20 min at 4 °C), the supernatant was discarded, and the pellet containing the intestinal brush border enzymes (alkaline phosphatase, aminopeptidase-N and maltase) dissolved in 1 mL of Tris-Mannitol.

Total alkaline protease activity was measured using azocasein (0.5%) as substrate in Tris-HCl 50 nmol $L^{-1}$ (pH = 9). One unit (U) of activity was defined as the nmoles of azo dye released per minute and per mL of tissue homogenate, and the absorbance read at λ = 366 nm. Trypsin activity was assayed using BAPNA (N-α-benzoyl-DL-arginine p-nitroanilide) as substrate. One unit of trypsin per mL (U) was defined as 1 μmol BAPNA hydrolyzed $min^{-1}$ $mL^{-1}$ of enzyme extract at λ = 407 nm (Holm et al., 1988). Chymotrypsin activity was quantified using BTEE (benzoyl tyrosine ethyl ester) as substrate and its activity (U) corresponded to the μmol BTEE hydrolyzed $min^{-1}$ $mL^{-1}$ of enzyme extract at λ = 256 nm (Worthington, 1991). Alpha-amylase activity was determined using 0.3% soluble starch as substrate (Métais and Bieth, 1968), and its activity (U) was defined as the amount of starch (mg) hydrolysed during 30 min per mL of tissue homogenate at λ = 580 nm. Bile salt-activated lipase activity was assayed for 30 min using p-nitrophenyl myristate as substrate. The reaction was stopped with a mixture of acetone: n-heptane (5:2), the extract centrifuged (2 min at 6,080 x *g* and 4 ºC) and the increase in absorbance of the supernatant read at λ = 405 nm. Lipase activity (U) was defined as the amount (nmol) of substrate hydrolyzed per min per mL of enzyme extract (Iijima et al., 1998). Pepsin activity (U) was defined as the nmol of tyrosine liberated per min per mL of tissue homogenate read at λ = 280 nm (Worthington, 1991).

Regarding intestinal digestive enzymes, alkaline phosphatase was quantified using 4-nitrophenyl phosphate (PNPP) as substrate. One unit (U) was defined as 1 μmol of pNP released $min^{-1}$ $mL^{-1}$ of brush border homogenate at λ = 407 nm (Gisbert et al., 2018). Aminopeptidase-N was determined using 80mM sodium phosphate buffer (pH = 7.0) and L-leucine p-nitroanilide as substrate (in 0.1 mM DMSO) (Maroux et al., 1973). One unit of enzyme activity (U) was defined as 1 μg nitroanilide released per min per mL of brush border homogenate at λ = 410 nm. Maltase activity was determined using d(+)-maltose as substrate in 100 mM sodium maleate buffer (pH = 6.0) (Dahkqvist, 1970). One unit of maltase (U) was defined as μmol of glucose liberated per min per mL of homogenate at λ = 420 nm. The assay of the cytosolic peptidase, LAP was performed on intestinal homogenates applying the method described by Nicholson and Kim (1975) which utilized L-alanine as substrate in 50 mM Tris-HCl buffer (pH = 8.0). One unit of enzyme activity (U) was defined as 1 nmol of the hydrolyzed substrate $min^{-1}$ $mL^{-1}$ of tissue homogenate at λ = 530 nm. Soluble protein of crude enzyme extracts was quantified by means of the Bradford's method (Bradford, 1976) using bovine serum albumin as standard. All enzymatic activities were measured at 25-26 ºC and expressed as specific activity defined as units per mg of protein (U mg protein$^{-1}$). All the assays were made in triplicate (methodological replicates) for each tank and the absorbance was read using a spectrophotometer (Tecan$^{TM}$ Infinite M200, Switzerland).

*2.7 Humoral immunity on skin mucus*

*2.7.1. Lysozyme, protease, antiprotease and total protein determination*

Lysozyme activity was measured according to the turbidimetric method described by Swain et al. (2007). Briefly, 20 µL of skin mucus were placed in flat-bottomed 96-well plates. To each well, 180 µL of freeze-dried *Micrococcus lysodeikticus* (0.2 mg mL$^{-1}$, Sigma-Aldrich) in 40 mM sodium phosphate (pH 6.2) was added as lysozyme substrate. As blanks of each sample, 20 µL of skin mucus were added to 180 µL of sodium phosphate buffer. The absorbance at λ = 450 nm was measured after 20 min at 35 ºC in a microplate reader (BMG Labtech). The amounts of lysozyme present in the samples were obtained from a standard curve made with hen egg white lysozyme (HEWL, Sigma) through serial dilutions in the above buffer. Skin mucus lysozyme values are expressed as U mL$^{-1}$ equivalent of HEWL activity.

Protease activity was quantified using the azocasein hydrolysis assay according to Guardiola et al. (2014). Aliquots of 100 µL of each mucus sample were incubated with 100 µL of 100 mM ammonium bicarbonate buffer containing 0.7% azocasein (Sigma-Aldrich) for 19 h at 30 ºC. The reaction was stopped by adding 4.6% trichloro acetic acid (TCA) and the mixture centrifuged (10,000 x *g*, 10 min). The supernatants were transferred to a 96-well plate in triplicate containing 100 µL well$^{-1}$ of 0.5 N NaOH. In both cases, the OD was read at λ = 450 nm using a plate reader. Skin mucus was replaced by trypsin (5 mg mL$^{-1}$, Sigma), as positive control (100% of protease activity), or by buffer, as negative controls (0 % of protease activity).

Total antiprotease activity was determined in skin mucus by its ability to inhibit trypsin activity (Hanif et al., 2004). Briefly, 10 µL of skin mucus were incubated (10 min, 22 ºC) with the same volume of standard trypsin solution (5 mg mL$^{-1}$) in a 96-well flat-bottomed plate. After adding a volume of 100 µL of 100 mM ammonium bicarbonate buffer and 125 µL of buffer containing 2% azocasein (Sigma), samples were incubated (2 h, 30 ºC) and, following the addition of 250 µL 10% TCA, were incubated again (30 min, 30 ºC). The mixture was then centrifuged (10,000 x *g*, 10 min) and the supernatant was transferred to a 96-well plate in triplicate, containing 100 µL well$^{-1}$ of 1 N NaOH before the OD was read at λ = 450 nm using a plate reader. For a positive control, the reaction buffer replaced mucus and trypsin, and for a negative control, the reaction buffer replaced the mucus. The antiprotease activity was expressed in terms of the percentage of trypsin inhibition according to the formula: % Trypsin inhibition = (Trypsin OD –Sample OD)/ Trypsin OD x 100.

Skin mucus protein concentration was determined by the dye binding method of Bradford (1976) using bovine serum albumin (BSA, Sigma-Aldrich) as the standard. Briefly, 2 mg mL$^{-1}$ solution of BSA was prepared and serial dilutions made with phosphate buffer saline (PBS Sigma-Aldrich) as standards. Dilutions of 5 µL of skin mucus and 15 µL of PBS were prepared. Then 250 µL of Bradford reagent (Sigma-Aldrich) was added to BSA and skin mucus dilutions and incubated at room temperature for 10 min. The absorbance of each sample was then read at λ = 595 nm and the results were taken and plotted onto the standard curve to obtain the total protein content of skin mucus. All spectrophotometry reads were conducted with a Varioskan 2.4.5, (Thermo Scientific, MA, USA ).

*2.8 Gut bacterial community DNA extraction and sequencing*

Total bacterial DNA was extracted and analysed from individual distal intestine content obtained from 5 fish per tank as previously reported in Parma et al. (2019). Afterwards, the V3–V4 hypervariable region of the 16S rRNA gene was amplified using the 341F and 785R primers (Klindworth et al., 2013) with added Illumina adapter overhang sequences and 2x KAPA HiFi HotStart ReadyMix (KAPA Biosystems). Briefly, the thermal cycle consisted of an initial denaturation at 95 °C for 3 min, 30 cycles of denaturation at 95 °C for 30 s, annealing at 55 °C for 30 s and extension at 72°C for 30 s, and a final extension step at 72 °C for 5 min. PCR reactions were cleaned up for sequencing by using Agencourt AMPure XP magnetic beads as recommended in the Illumina protocol "16S Metagenomic Sequencing Library Preparation" for the MiSeq system, and as used in several other publications (Biagi et al., 2018; Soverini et al., 2016). Sequencing was performed on Illumina MiSeq platform using a 2 x 250 bp paired-end protocol according to the manufacturer's instructions (Illumina, San Diego, CA). The sequencing process resulted in a total of 1,553,593 high quality reads that were processed using the QIIME 2 pipeline (Bolyen et al., 2019). After length (minimum/maximum = 250/550 bp) and quality filtering with default parameters, reads were cleaned using DADA2 (Callahan et al., 2016) and clustered into OTUs at a 0.99 similarity threshold using VSEARCH (Rognes et al., 2016). Assignment was carried out by using the RDP classifier against Silva database (Quast et al., 2013).

*2.9 Metabolic parameters in plasma*

The levels of glucose (GLU), urea, creatine, uric acid, total bilirubin, bile acid, amylase, lipase, cholesterol (CHOL), triglycerides (TRIG), total protein (TP), albumin (ALB), aspartate aminotransferase (AST), alanine transaminase (ALT), alkaline phosphatase (ALP), gamma-glutamyl transferase (GGT), creatine kinase (CK), lactate dehydrogenase (LDH), calcium ($Ca^{+2}$), phosphorus (P), potassium ($K^+$) sodium ($Na^+$), iron (Fe), chloride (Cl), magnesium (Mg), unsaturated iron binding capacity (UIBC), total iron binding capacity (TIBC) and cortisol were determined in the plasma using samples of 500 μL on an automated analyser (AU 400; Beckman Coulter) according to the manufacturer's instructions. The ALB/globulin (GLOB), Na/K ratio and Ca x P were calculated.

*2.10 Statistical analysis*

All data are presented as mean ± standard deviation (SD). A tank was used as the experimental unit for analysing growth performance and a pool of five and three sampled fish were considered the experimental unit for analysing carcass composition and enzyme activity respectively. Individual fish were used for analysing plasma biochemistry and mucus stress parameters. Data of growth performance, nutritional indices, enzyme activity, plasma and skin mucus parameters were analysed by a two-way analysis of variance (ANOVA) and in case of significance ($p \leq 0.05$) Tukey's post hoc test was performed. The normality and/or homogeneity of variance assumptions were validated for all data preceding ANOVA. The R packages "Stats" and "Vegan" were used to perform gut microbiota statistical analysis. In particular, to compare the microbiota structure among different groups for alpha and beta-diversity, Wilcoxon rank-sum test was used while the PCoA was tested using a permutation test with pseudo-F ratios (function "Adonis" in the "Vegan" package). Alpha diversity of the different ecosystems was computed using Hill numbers (Hill, 1973; Chao et al., 2014). Beta diversity was estimated using both weighted and unweighted UniFrac metrics. Statistical analyses were performed using GraphPad Prism 6.0 for Windows (Graph Pad Software, San Diego, CA, USA) and RStudio interface for R (https://www.r-project.org). The differences among treatments were considered significant at $p \leq 0.05$.

## 3. Results

*3.1 Growth*

Results on growth performance parameters are summarised in Table 2. No significant effects on growth (FBW, weight gain and SGR) were detected between LD and HD groups for both dietary treatments ($p > 0.05$). However, fish fed FM30/FO15 displayed higher FBW, weight gain and SGR values compared to the FM10/FO3 group ($p < 0.05$). Values of FI were lower in HD compared to LD (density effect $p = 0.002$) with more marked differences in FM30/FO15 then FM10/FO3, whereas no significant diet effect on FI was detected ($p > 0.05$). No significant effect of density on FCR was observed ($p > 0.05$), while the FM10/FO3 group showed higher FCR values, followed by FM30/FO15. Survival rates were lower in the LD group ($p < 0.05$).

Data on body composition and nutritional indices are shown in Table 3. Whole body composition values were not significantly influenced by different fish density ($p > 0.05$), while lipid content was lower in fish fed the FM10/FO3 diet compared to the FM30/FO15 group ($p < 0.05$); however, ash and moisture levels were higher in FM10/FO3 than FM30/FO15 fish ($p < 0.05$). No significant effects of fish density on PER, GPE, GLE and LER were detected ($p > 0.05$); however, fish fed FM10/FO3 displayed lower PER, GPE, GLE and LER compared to FM30/FO15 ($p < 0.05$).

*3.2 Digestive enzyme activity*

Data on specific activity of gastric, pancreatic and intestinal digestive enzymes are shown in Table 4. The activities of both pancreatic (trypsin, chymotrypsin, total alkaline proteases, amylase and bile salt-activated lipase) and gastric (pepsin) enzymes were not significantly affected by the rearing density nor the diet ($p > 0.05$); with the exception of trypsin, which was slightly affected by the diet composition ($p = 0.053$) with lower values recorded in fish fed the FM10/FO3 diet compared to those fed the FM30/FO15 diet. Regarding intestinal brush border enzymes measured in the anterior segment of the intestine, aminopeptidase-N and maltase activities were not significantly affected by the diet nor rearing density ($p > 0.05$), while phosphatase alkaline and LAP were slightly ($p < 0.1$) lower in FM10/FO3 than FM30/FO15. The activity of LAP was significantly higher at HD compared to LD for both dietary treatments ($p < 0.05$). Concerning the intestinal enzymes measured in the posterior region of the intestine, aminopeptidase and LAP were significantly affected by the rearing density with lower values recorded at HD in comparison to those recorded in fish kept at LD ($p < 0.05$). Diet significantly affected aminopeptidase-N and maltase activities which were significantly lower in sea bream fed the FM10/FO3 diet ($p < 0.05$). No significant effects of both diets and tested densities were detected in the phosphatase alkaline activities in the posterior intestine ($p > 0.05$).

*3.3 Plasma biochemistry*

The results of plasma parameters are shown in Table 5. No significant effect ($p > 0.05$) of density on plasma parameters was detected under both feeding regimes. Concerning the effect of diet on plasmatic parameters like urea, lipase, UIBC, A/G, TIBC, $Na^+$, $K^+$, $Cl^-$, these were higher in fish from the FM10/FO3 group compared to those from the FM30/FO15 group ($p < 0.05$), while creatine, $Ca^{2+}$, Mg, CHOL, TP, ALB and $Na^+/K^+$ were lower in FM10/FO3 compared to FM30/FO15 fish ($p < 0.05$). No significant differences related to density and feeding regimes for GLU, uric acid, creatine, total bilirubin, AST, ALT, ALP, amylase, GGT, CK, LDH, P, TRIG, Bile acid, CaxP, Fe and cortisol were detected among experimental groups ($p > 0.05$).

*3.4 Skin mucus non-specific immune biomarkers*

Results of skin mucus lysozyme, protease, antiprotease and total proteins are presented in Figure 1 (A-D). Lysozyme activity was slightly affected by the rearing density (density effect $p = 0.04$) with higher values recorded under LD rearing conditions. Specifically, lysozyme was significantly higher in fish fed FM30/FO15 at LD rearing conditions compared to those fed FM10/FO3 and reared at HD (Fig 1A; $p < 0.05$). Protease was significantly reduced under fish fed FM10/FO3 (diet effect $p = 0.0006$), while no significant effect of rearing density was detected ($p > 0.05$). Specifically, protease activity in skin mucus was significantly higher in fish fed the FM30/FO15 diet at both rearing densities compared to those fed FM10/FO3 and reared at LD (Fig 1B; $p < 0.05$). No significant effect of density or diet were detected in antiprotease activity and total proteins of skin mucus from fish belonging to the different experimental groups (Fig. 1, C-D; $p > 0.05$).

*3.5 Gut bacterial community profiles*

Taxonomic characterisation of the gut bacterial community at different phylogenetic levels is represented in Figure 2: phylum in panel (A) and family in panel (B) and in Supplementary Table 1. At phylum level, the most abundant taxa were Firmicutes, Actinobacteria and Proteobacteria. In addition, the families most represented, all belonging to Firmicutes phylum, were *Lactobacillaceae* (FM30/FO15$_{HD}$: 77.9% ± 16.1%; FM30/FO15$_{LD}$: 86.5% ± 4.4%; FM10/FO3$_{HD}$: 61.3% ± 12.4%; FM10/FO3$_{LD}$: 67.6% ± 12.2%), *Streptococcaceae* (FM30/FO15$_{HD}$: 2.0% ± 1.5%; FM30/FO15$_{LD}$: 1.3% ± 1.4%; FM10/FO3$_{HD}$: 4.1 % ± 3.7%; FM10/FO3$_{LD}$: 3.2% ± 2.3%) and *Staphylococcaceae* (FM30/FO15$_{HD}$: 1.4 % ± 1.0 %; FM30/FO15$_{LD}$: 0.9 % ± 0.4 %; FM10/FO3$_{HD}$: 0.6% ± 1.3%; FM10/FO3$_{LD}$: 0.3% ± 0.5%). No significant differences (Wilcoxon test $p > 0.05$, FDR correction) among groups at phylum level were detected between specimens fed with the same diet but in different rearing density condition. On the other hand, significant differences in several families such as *Staphylococcaceae* were observed, values that were higher in the FM30/FO15$_{HD}$ group than in FM10/FO3$_{HD}$ group ($p < 0.05$, Wilcoxon rank-sum test), and *Streptococcaceae*, higher in FM10/FO3$_{HD}$ group compared to FM30/FO15$_{HD}$ group ($p < 0.05$). Moreover, at LD, both diets determined a significant difference in the abundance of *Lactobacillaceae* and *Staphylococcaceae*, both higher in FM30/FO15$_{LD}$ group compared to FM10/FO3$_{LD}$ ($p < 0.05$, Wilcoxon rank-sum test) (Figure 2 C).

The biodiversity among microbiota from fish fed different diets and kept at different stocking densities, expressed using Hill numbers of different magnitudes (from q = 0 to q = 2), is represented in panel A of Figure 3. For all the q value magnitude, diet FM10/FO3 is characterised by a more even distribution of bacterial species characteristic that is strengthened going from order q 0 to order q 2. According to the results, diet FM10/FO3 was more effective in the maintenance of a greater biodiversity in the sea bream gut ecosystem. Furthermore, it is interesting to notice that for a q = 0, diet FM30/FO15 showed a number of species comparable to diet FM10/FO3, shifting to a significantly more uneven ecosystem ($p < 0.05$, t-test) increasing the weight of the microbial core (q values of 1 and 2, respectively). These results also showed that the response to rearing conditions shifted depending on the fishes feeding regimen: diet FM10/FO3 maintained steady the biodiversity of the gut microbiota between HD and LD ($p$ value $> 0.05$; t-test). On the other hand, diet FM30/FO15 was not able to maintain the evenness of the community, as highlighted in the q value of 2, in which the FM30/FO15$_{HD}$ group showed a significantly reduced biodiversity when compared to the other groups (p value $< 0.05$, t-test). To assess whether these different treatments could influence the gut bacterial ecosystem, a multivariate analysis was performed. In both Principal Coordinates Analysis (PCoA) graphs obtained using both weighted UniFrac metric (Figure 3 B) and unweighted UniFrac metric (Figure 3 C) a significant separation was observed between the different groups in the two-dimensional space (Adonis $p < 0.01$), except for FM30/FO15$_{HD}$ *vs* FM30/FO15$_{LD}$ which did not show a significant ($p > 0.05$) separation.

**Discussion**

Several studies have investigated the effect of high rearing density on growth, physiological responses and health in gilthead sea bream; however, studies concerning the possible interaction between rearing density and low FM FO-based diets have been less explored. In the present study, fish reared at high density (14.5-36/44 kg m$^{-3}$, initial and final density, respectively) within each FM and FO dietary levels showed similar performance in terms of growth and feed utilisation in comparison to those reared at low density (4.8-12/15 kg m$^{-3}$). The results of the present study during the on-growing phase (96-318g) go beyond the maximum density tested (20-31 kg m$^{-3}$) by Araújo-Luna et al. (2018) for gilthead sea bream at similar size (268-435 g). The authors did not find any negative effects of high rearing density on SGR even if a significant linear relationship between FCR and increasing stocking densities was observed. Indeed, the results of the present study are consistent with a previous observation reported on juveniles (22-85 g) in which high density up to 40.8 kg m$^{-3}$ did not negatively affect growth (Montero et al., 1999). However, more recently, Diogenes et al. (2019) found that rearing density up to 57 kg m$^{-3}$ impaired FI, growth and FCR in sea bream juveniles (12-58g). The authors suggested that 40 kg m$^{-3}$ could be near the maximum tolerable stocking density for gilthead sea bream of the weight range tested. This seems in agreement also for the size tested in the present study; even if high density had no negative effect on the overall growth and feed utilisation, high density significantly ($p = 0.002$) reduced FI.

Interestingly, this effect was mainly reported in high FM and FO dietary level and this could be a consequence of the higher final stocking density obtained under this treatment (44 vs 36 kg m$^{-3}$, FM30/FO15, FM10/FM3, respectively) or be due to the fact that density could have increased feeding competition only in a potentially more palatable and digestible diet. The differences observed in growth performance between diets were mainly related to a lower feed utilisation occurring in FM10/FO3; however it should be taken into account that the growth performance achieved in the present trial under both diets is in line with those found in literature for similar dietary formulation and that the sole comparison between the two diets was not the purpose of the present study.

Stress conditions can disrupt the endocrine system and affect some physiological functions such as digestive capacity (Trenzado et al., 2018). Few studies have evaluated the effect of stocking density with a dietary interaction on digestive enzyme activity at the on-growing stage in fish species (Wong et al., 2013). In the present study rearing density did not affect pancreatic digestive enzyme specific activities under both dietary treatments. Similarly, protease, lipase and amylase activities were not affected by rearing density in gilthead sea bream fed increasing dietary tryptophan level with alternative vegetable protein sources (Diogenes et al., 2019) or in Nile tilapia (*Oreochromis niloticus*) fed dietary live and heat-inactive baker's yeast in vegetable-meal based diet (Ran et al., 2016). Contrarily, Trenzado et al. (2018) studying the interaction between stocking density and dietary lipid content in rainbow trout (*Oncorhynchus mykiss*) found that stocking density inhibited the adaptive response of lipase activity and enhanced the protease activity inhibition due to higher dietary lipid content. Compared to the pancreatic enzyme activity, in the present study, density seemed to slightly affect the proteolytic enzyme activity measured in the intestinal brush border of enterocytes. In particular, LAP activity measured in the brush border of the anterior intestine tended to increase at high density while aminopeptidase and LAP activity in the posterior intestine was slightly reduced at high density in particular in the low FM-FO diet. The alkaline phosphatase of the intestinal brush border is used as a marker of intestinal integrity and among its functions was found to keep gastrointestinal inflammation under control (Lalles et al., 2019, Messina et al., 2019). In addition, Nile tilapia reared at higher density displayed higher alkaline phosphatase activity, possibly in line with higher pathogenic stressors at high rearing density (Ran et al., 2016). In the present study, the absence of differences in the alkaline phosphatase activity suggested no major functional changes in the integrity of the intestine under different rearing density in both dietary treatments. The evaluation of several plasma biochemical parameters is considered a valuable approach for assessing the suitability of feeding practices, metabolic disorders, rearing conditions and presence of acute or chronic stressors (Peres et al., 2013; Guardiola et al., 2018). No significant effect of stocking density on any of the twenty-seven different plasma parameters measured was detected under both dietary treatments. It is commonly accepted that high stocking density generally leads to increased plasma cortisol levels in different fish species, enhancing metabolic rate and compromising energy availability for several physiological processes such as growth (Ashley, 2007). However, an opposite cortisol response to stocking density has been also observed in some fish species suggesting that cortisol response to stocking density is species-dependent and related to the gregarious behaviour of the species at a specific stage of life (De las Heras et al., 2015; Millán-Cubillo et al., 2016). Previous study of juveniles and adult sea bream held at high stocking density, giving rise to chronic stress, showed significantly higher levels of plasma cortisol than those held at low density, suggesting the incapacity of this species to reach adaptation under chronic high rearing density conditions (Montero et al., 1999; Sangia-Alvarellos et al., 2005). In accordance, TP, CHOL, TRIG were also found to be reduced at high stocking density as a consequence of increased energy demand under stressful conditions and possibly mediated by increased plasma cortisol (Diogenes et al., 2019). As also reported for Senegal sole (*Solea senegalensis*) by Azeredo et al. (2019) the fact that fish held at high density did not show higher plasma cortisol than their low-density counterparts might be related to negative feedback mechanisms established in the HPI axis, as a strategy of chronically stressed animals to attenuate an exacerbated stress response (Bonga, 1997; Mommsen et al., 1999). In addition, the absence of effects of rearing density on GLU, CHOL, TP and TRIG, suggests that the differences in rearing density were not able to alter the metabolic processes related to growth and feed utilisation. Non-specific plasma enzymes, such as AST, GGT, ALP, CK and LDH are considered useful indicators of the health status and their elevated plasma level may indicate specific tissue damage of several organs including liver, muscle, spleen and kidney related to pathological processes, toxic chemical exposure, or traumatic conditions or hypoxia, whereas specific references for this species and age are few (Peres et al., 2013; Guardiola et al., 2018). Values of AST, CK, GGT and LDH were found in the lower part of the range proposed by Peres et al. (2013) for healthy juvenile sea bream (70 g) fed FM-based diet at low rearing density (3-5 kg $m^{-3}$) and in line with those found by Guardiola et al. (2018) during a feeding trial in sea bream of similar size. Levels of ALP were higher than values previously found by Peres et al. (2013) and Guardiola et al. (2018), a difference which can be related to FI since this enzyme is involved in the absorption and transport of lipid and carbohydrates from the intestine, and its intestinal activities are positively correlated with food ingestion and growth rate (Lemieux et al., 1999; Lalles et al., 2019). The values of plasma electrolytes provided in the trial were comparable with the values reported in sea bream (Peres et al., 2013; Guardiola et al., 2018) and sobaity sea bream (*Sparidentex hasta*) (Hekmatpoure et al., 2019). Plasma electrolytes are indicators of the secondary phase of stress response in fish, providing an indirect indication of altered plasma cortisol levels; in particular plasma phosphorus and calcium levels were found to be sensitive to fish stocking density (Hrubec et al., 2000) while potassium levels are accepted as a general indicator of stress in fish (Guardiola et al., 2018).

Evaluation of skin mucosal immunity has been proposed recently as a promising alternative stress assessment in fish species after stressful conditions including crowding or transportation, whereas data of specific mucosal component in response to different stressors are still scarce (Guardiola et al 2016; Sanahuja et al., 2019). Enzymes in the epidermal mucus such as lysozyme, protease and antiprotease play an important role in humoral and skin mucus defence acting directly on a pathogen, or activating and enhancing the production of various immunological components of fish subjected to stressful situations (Esteban, 2012; Guardiola et al., 2016). The present results indicate different effects of treatments on specific skin mucus components, lysozyme being slightly reduced by high rearing density while protease was mainly reduced by low FM-FO diets. Both enzymes have been shown to be modulated either by diet or environmental conditions in sea bream. Most studies have shown the possibility of increasing lysozyme activity of skin mucus by dietary additives, such as selenium nanoparticles, *Moringa oleifera* leaves or probiotics; but crowding conditions at 20 kg m$^{-3}$ for 30 days has also been reported to lead to an increase in lysozyme gene expression in sea bream skin mucus (Cordero et al., 2016; Mansour et al., 2018; Dawood et al., 2019). Concerning protease activity, Guardiola et al. (2016) found a significant increase in this activity after 24 and 48 h of acute 50 kg m$^{-3}$ crowding stress. However, in the same study a reduction in the protease activity was also found after 48 h. The effect of protease activity under chronic stressful conditions has been poorly investigated. Easy et al. (2010) studied the skin mucus components following short- and long-term handling stress in Atlantic salmon (*Salmo salar*), and no correspondence between skin mucus component and plasma cortisol level in long-term stress was observed, suggesting that the activation of mucus proteases may have been triggered by short-term elevated cortisol levels or that skin mucus protease activation could result from physical disturbances such as abrasion due to netting or overcrowding. More studies are needed to understand the role played by skin mucus on stress in fishes.

Although the study of the gut microbiota by next-generation sequencing (NGS) has already been conducted in this species under different feeding treatment, no information concerning the effects of rearing density on gut microbiota is available. According to our findings, the gut bacterial community is dominated by Firmicutes (69.9-92.2%), followed by Actinobacteria and Proteobacteria. The dominance of Firmicutes we observed is in general agreement with the previous NGS-based survey of the gut bacterial community in sea bream and other marine or freshwater species fed similar aquafeed ingredients employed in the present study (FM, soy-derivates, corn glutens, wheat gluten and wheat meal) (Parma et al., 2016, Rimoldi et al., 2018a, 2018b; Parma et al., 2019). However, our data differ from previous findings concerning the gut bacterial community of gilthead sea bream and other Mediterranean fish species which displayed a dominance of Proteobacteria and detected Firmicutes as the subdominant component (Carda-Diéguez et al., 2014; Gatesoupe et al., 2016, Piazzon et al., 2017). These works characterised the mucosa-adherent gut microbiota, which could differ from the microbiota of the intestinal lumen (Ringo et al., 2018). In this context, a recent comparison between mucosa-adherent gut microbiota and intestinal lumen gut microbiota in sea bream highlighted the dominance of Proteobacteria in the gut mucosa while Firmicutes dominated the intestinal lumen in the same specimens (unpublished data). In addition, other studies revealed that the differences in abundance between Firmicutes and Proteobacteria could also have been related to the dietary composition. In rainbow trout, the presence of Proteobacteria was favoured by an animal protein-based diet while the inclusion of at least 25% of plant proteins in the diet favoured the presence of Firmicutes (Rimoldi et al., 2018b).

At the family level, the gut bacterial community of the present study was widely dominated by *Lactobacillaceae* ranging from 61.3 to 86.5 %. The presence and the role of *Lactobacillaceae* and other lactic acid bacteria (LAB) in fish species is still controversial (Ringo et al., 2018). Several studies have associated a high LAB abundance with a high inclusion level of dietary plant ingredients or functional additives in sea bream (Parma et al., 2016; Rimoldi et al., 2018a) or other marine fish species (Apper et al., 2016; Rimoldi et al., 2018b; Parma et al., 2019). However, some studies found a reduction in LAB relative abundance when high FM replacement was also associated with a decrease in performance (Estruch et al., 2015; Miao et al., 2018), while others found a higher abundance of LAB in relation to vegetable protein associated with impaired gut health (Gajardo et al., 2017). The results of the present study reinforce previous observation that the dominance of *Lactobacillaceae* mainly *Lactobacillus* could be considered a valid indicator of optimal gut health condition in sea bream.

No significant differences related to rearing density of any specific component within each diet at phylum level were detected (Wilcoxon ran-sum test, $p > 0.05$, FDR correction). However, different responses of the intestinal gut microbial composition in relation to dietary treatment under high and low rearing density were detected as also highlighted by weighted and unweighted UniFrac PCoA. In particular, no significant separation was found between densities when fish were fed high FM-FO level, while under low FM-FO diet density had a significant effect. Focusing on specific components of the gut bacterial community, the results indicated that under high rearing density high FM-FO level led to a significant increase in *Staphylococcaceae* and a reduction in *Streptococcacee* abundances compared to low FM-FO diet, while under low rearing density *Lactobacillaceae* were less abundant in low FM-FO diet than high FM-FO diet. Although no significant differences were detected, high rearing density seems to reduce the amount of *Lactobacillaceae* (mainly *Lactobacillus spp*) within each dietary treatment (Supplementary Table 1). No studies are available to compare the effect of rearing density on specific gut microbial components in fish. In the present study, no evident signs of stress induced by high rearing density were detected by results of performance, plasma and skin mucus parameters; however, *Lactobacillaceae* may be highly sensitive in relation to environmental stressors in fish and may deserve further attention for future studies.

Analysis of biodiversity of the microbial community has highlighted a different response to the feeding regimes, showing a general higher biodiversity in fish fed diets containing higher vegetable ingredients. This is in general agreement with previous findings detecting feeding habit as a key factor influencing fish gut microbial diversity and observing an increasing trend in diversity following the order of carnivores, omnivores and herbivores (Wang et al., 2018). In addition, a significant increase in α-diversity indices at increasing FM replacement with vegetal ingredients was observed in carnivorous fish species (Desai et al 2012; Miao et al., 2018). Concerning the interaction between diet and rearing density, a low FM-FO diet maintained steady the biodiversity of the ecosystem between low and high-density conditions while fish fed high FM-FO level showed a significantly reduced biodiversity at high rearing density when compared to the other groups. It has been suggested that in fish, reduction in diversity leads to reduced competition for opportunistic or invading pathogens which may enter the gastrointestinal tract of fish via feed or water (Apper et al., 2016). In several fish species, α-diversity was not found to be affected by dietary vegetal ingredients (Apper et al., 2016; Parma et al., 2016; Rimoldi et al., 2018b), by the interaction between diet and rearing density (Wong et al., 2013) or by stocking density (Du et al., 2019). Also in pigs, stocking density did not significantly affect biodiversity indices of gut microbiota (Li et al., 2017). Interestingly, recent findings in the African cichlid *Astatotilapia burtoni* highlighted that fish which experienced stressful conditions induced by subordinate social rank displayed a reduced faecal microbial community α-diversity (Singh et al., 2019). Also in captive mice and in wild red squirrels (*Tamiasciurus hudsonicus*) socially mediated stress affected the intestinal microbiota leading to a reduction in microbial diversity and richness (Bailey et al., 2011; Stothart et al., 2016). The reduction of biodiversity observed in the present study only under the high FM-FO level could be correlated to increased feeding competition only when a potentially more palatable high FM-FO diet is offered. Another explanation may be associated with the lower feed intake observed under high rearing density when fed high FM-FO level, or a combination of both factors: feeding competition and feed intake. Recently, in perch (*Perca fluviatilis*) Zha et al. (2018) found that gut microbial diversity responded to predation stress and food ration with a reduction in diversity due to the presence of a predator and a reduced feed ration. The authors suggested that a high ration of food favours bacteria that are quick colonizers and fast growers while at lower food rations bacteria that are good competitors would be favoured. In addition, the fact that in our study the reduction in gut microbial diversity was not supported by evident altered physiological signs of stress could indicate a high sensitivity of the gut microbial community structure to food competition, or to other social interaction induced by rearing density. Thus, the analysis of gut microbial community diversity could represent a valuable tool to assess social stress conditions for future studies related to feeding behaviour and feeding competition.

**Conclusion**

In conclusion, the different rearing densities tested in this trial had no major effects on overall performances and feed efficiency of gilthead sea bream reared at high or low fish meal and fish oil dietary level. However, rearing density reduced feed intake in fish fed high fish meal and fish oil dietary level. Results of digestive enzyme activities indicated a comparable digestive efficiency among rearing densities and within dietary treatment even if intestinal brush boarder enzymes such as LAP and aminopeptidase seems to be more influenced by stocking density compared with other (gastric and pancreatic) enzymes. Plasma parameters related to nutritional and physiological conditions were not affected by rearing densities, indicating that sea bream can well cope with high rearing density up to 36-44 kg $m^{-3}$ and that a high level of vegetable dietary ingredients does not amplify the potential stressful effects of rearing density. A similar observation was achieved through the study of skin mucosal immunity; however in this case lysozyme was slightly reduced at high density. For the first time the effect of rearing density on gut bacterial community of this species was studied. Different responses in relation to dietary treatment under high and low rearing density were detected. Low FM-FO diet maintained steady the biodiversity of gut bacterial community between low and high rearing density while fish fed high FM-FO level showed a significantly reduced biodiversity at high rearing density possibly indicating higher social stress conditions related to feeding competition under this treatment. According to the results, it seems feasible to rear gilthead sea bream at the on-growing phase at a density up to 36-44 kg m$^{-3}$ with low or high FM-FO diet without negatively affecting growth, feed efficiency, welfare condition and gut microbial community.

**Acknowledgment**

This research was supported by ERC (European Research Council) in MedAID project (Mediterranean Aquaculture Integrated Development), Call H2020-SFS-2016-2017 (Sustainable Food Security – Resilient and resource-efficient value chains), Grant Agreement n. 727315. Analyses of digestive enzymes conducted at IRTA were partially supported by the project ADIPOQUIZ (RTI2018-095653-R-I00) funded by the Ministerio de Ciencia, Innovación y Universidades (Spain). The authors would like to thank Gillian Forlivesi Heywood for English language editing and Stefano Porcelli for the technical contribution in fish rearing and laboratory analysis.

Deguara, S., Jauncey, K., Agius, C., 2003. Enzyme activities and pH variations in the digestive tract of gilthead sea bream. J. Fish Biol., 62(5), 1033-1043.

Desai, A.R., Links, M.G., Collins, S.A., Mansfield, G.S., Drew, M.D., Van Kessel, A.G., Hill, J.E., 2012. Effects of plant-based diets on the distal gut microbiome of rainbow trout (*Oncorhynchus mykiss*). Aquaculture 350, 134-142.

Diógenes, A.F., Teixeira, C., Almeida, E., Skrzynska, A., Costas, B., Oliva-Teles, A., Peres, H., 2019. Effects of dietary tryptophan and chronic stress in gilthead seabream (*Sparus aurata*) juveniles fed corn distillers dried grains with solubles (DDGS) based diets. Aquaculture 498, 396-404.

Du, F., Li, Y., Tang, Y., Su, S., Yu, J., Yu, F., Li, J., Li, H., Wang, M., Xu, P., 2019. Response of the gut microbiome of *Megalobrama amblycephala* to crowding stress. Aquaculture 500, 586-596.

Easy, R.H., Ross, N.W., 2010. Changes in Atlantic salmon *Salmo salar* mucus components following short- and long-term handling stress. J. Fish Biol, 77, 1616-1631.

Ellis, T., North, B., Scott, A.P., Bromage, N.R., Porter, M., Gadd, D., 2002. Review paper: the relationships between stocking density and welfare in farmed rainbow trout. J. Fish. Biol. 61:493–531.

Esteban, M. A., 2012. An overview of the immunological defenses in fish skin, ISRN Immunol. 1–29 https://doi.org/10.5402/2012/853470.

Estruch, G., Collado, M.C., Peñaranda, D.S., Tomás Vidal, A., Jover Cerdá, M., Pérez Martínez, G., Martinez-Llorens, S., 2015. Impact of fishmeal replacement in diets for gilthead sea bream *Sparus aurata* on the gastrointestinal microbiota determined by pyrosequencing the 16S rRNA gene. PLoS One 10, e0136389.
36

**Table 1.** Ingredients and proximate composition of the experimental diets

|  | FM30/FO15 | FM10/FO3 |
|---|---|---|
| *Ingredients, % of the diet* | | |
| Fish meal (LT70) | 30.0 | 10.0 |
| Soybean meal 48 | 9.0 | 9.0 |
| Soy protein concentrate | 10.0 | 20.5 |
| Wheat gluten | 5.0 | 10.2 |
| Corn gluten | 10.0 | 15.0 |
| Wheat meal | 9.7 | 7.3 |
| Rapeseed meal | 5.0 | 4.0 |
| Sunflower meal | 5.0 | 4.0 |
| Fish oil | 15.0 | 3.0 |
| Rapeseed oil | 0 | 13.0 |
| Vit/Min premix[1] | 1.0 | 1.0 |
| Antioxidant powder (Paramega) | 0.2 | 0.2 |
| Sodium propionate | 0.1 | 0.1 |
| MCP | | 2.0 |
| Lysine | - | 0.3 |
| Methionine | - | 0.1 |
| L-Tryptophan | | 0.3 |
| *Proximate composition, % on a wet weight basis* | | |
| Moisture | 5.83 | 4.9 |
| Protein | 46.3 | 44.7 |
| Lipid | 17.2 | 17.8 |
| Ash | 8.2 | 6.4 |
| Gross energy cal g$^{-1}$ | 4945.7 | 4823.6 |

[1]Vitamins and mineral premix (IU or mg kg$^{-1}$ diet; Invivo NSA,: Portugal); DL-alpha tocopherol acetate, 200 mg; sodium menadione bisulphate, 10 mg; retinyl acetate, 16650 IU; DL-cholecalciferol, 2000 IU; thiamine, 25 mg; riboflavin, 25 mg; pyridoxine, 25 mg; cyanocobalamin, 0.1 mg; niacin, 150 mg; folic acid, 15 mg; L-ascorbic acid monophosphate, 750 mg; inositol, 500 mg; biotin, 0.75 mg; calcium panthotenate, 100 mg; choline chloride, 1000 mg, betaine, 500 mg; copper sulphate heptahydrate, 25 mg; ferric sulphate monohydrate, 100 mg; potassium iodide, 2 mg; manganese sulphate monohydrate, 100 mg; sodium selenite, 0.05 mg; zinc sulphate monohydrate, 200 mg
MCP: monocalcium phosphate

**Table 2.** Growth performance of gilthead sea bream reared at low and high stocking density and fed the experimental diets over 98 days.

|  | *Experimental diets* | | | | | *P value* | |
|---|---|---|---|---|---|---|---|
|  | FM30/FO15 | | FM10/FO3 | | *Density* | *Diet* | *Inter* |
|  | LD | HD | LD | HD | | | |
| Initial density kg m$^{-3}$ | 4.8±0.1$^a$ | 14.5±0.6$^b$ | 4.8±0.1$^a$ | 14.3±0.1$^b$ | *<0.0001* | *0.7078* | *0.7078* |
| Final density kg m$^{-3}$ | 15.2±0.5$^b$ | 43.6±0.5$^d$ | 12.1±1.3$^a$ | 35.9±0.5$^c$ | *<0.0001* | *<0.0001* | *0.0011* |
| IBW(g) | 96.1±1.1 | 96.4±3.7 | 96.6±2.6 | 95.5±0.8 | *0.768* | *0.878* | *0.630* |
| FBW(g) | 317.8±5.6$^b$ | 292.5±3.9$^b$ | 253.1±27.2$^a$ | 246.2±2.8$^a$ | *0.084* | *0.0001* | *0.292* |
| Weight gain (g) | 221.7±5.4$^b$ | 196.2±0.5$^b$ | 156.5±25.3$^a$ | 150.7±3.0$^a$ | *0.071* | *0.0001* | *0.224* |
| SGR | 1.22±0.02$^b$ | 1.13±0.03$^b$ | 0.98±0.09$^a$ | 0.97±0.02$^a$ | *0.127* | *0.0001* | *0.248* |
| FI | 15.6±0.19$^b$ | 14.6±0.21$^a$ | 15.4±0.64$^{ab}$ | 14.5±0.03$^a$ | *0.002* | *0.506* | *0.818* |
| FCR | 1.43±0.02$^a$ | 1.42±0.01$^a$ | 1.70±0.21$^b$ | 1.61±0.02$^{ab}$ | *0.433* | *0.005* | *0.495* |
| Survival % | 95.8±1.4$^a$ | 99.4±0.5$^b$ | 95.8±1.4$^a$ | 97.2±0.5$^{ab}$ | *0.004* | *0.111* | *0.111* |

Data are given as the mean (n=3) ± SD. In each line, different superscript letters indicate significant differences among treatments ($P \leq 0.05$). FM30/FO15 = 300g kg$^{-1}$ fishmeal (FM), 150 g kg$^{-1}$ fish oil (FO); FM10/FO3 = 100g kg$^{-1}$ FM; 30g kg$^{-1}$ FO. LD, low rearing density; HD, high rearing density.
IBW = Initial body weight.
FBW = Final body weight.
SGR = Specific growth rate (% day$^{-1}$) = 100 * (ln FBW- ln IBW) / days.
ABW = average body weight = (IBW + FBW)/2.
FI= Feed intake (g kg ABW$^{-1}$ day$^{-1}$) = ((1000*total ingestion)/(ABW))/days)).
FCR = feed conversion rate = feed intake (g) /weight gain (g)

**Table 3**. Body composition and nutritional indices of gilthead sea bream reared at low and high stocking density and fed the experimental diets over 98 days.

| | Experimental diets | | | | P-value | | |
| --- | --- | --- | --- | --- | --- | --- | --- |
| | FM30/FO15 | | FM10/FO3 | | | | |
| | LD | HD | LD | HD | *Density* | *Diet* | *Inter.* |
| Whole body composition, % | | | | | | | |
| Protein | 17.0 ± 0.5 | 17.2 ± 0.1 | 17.0 ± 0.0 | 16.9 ± 0.1 | *0.835* | *0.333* | *0.358* |
| Lipid | 21.4 ± 2.5$^b$ | 19.5 ± 1.5$^{ab}$ | 16.6 ± 0.7$^a$ | 17.0 ± 0.8$^a$ | *0.451* | *0.003* | *0.233* |
| Ash | 3.43 ± 0.11 | 3.57 ± 0.25 | 3.88 ± 0.08 | 3.83 ± 0.21 | *0.662* | *0.008* | *0.37* |
| Moisture | 58.0 ± 0.49 | 58.7 ± 0.7 | 59.5 ± 0.8 | 60.3 ± 0.9 | *0.206* | *0.024* | *0.949* |
| Nutritional indices | | | | | | | |
| PER | 1.51 ± 0.02 | 1.52 ± 0.01 | 1.32 ± 0.16 | 1.39 ± 0.02 | *0.443* | *0.009* | *0.567* |
| GPE | 25.8 ± 0.88 | 26.4 ± 0.38 | 22.6 ± 2.74 | 23.4 ± 0.20 | *0.455* | *0.006* | *0.879* |
| GLE | 101 ±14.8$^b$ | 91.7 ± 9.0$^b$ | 60.9 ± 9.4$^a$ | 66.2 ± 4.6$^a$ | *0.768* | *0.000* | *0.253* |
| LER | 4.08 ± 0.05$^b$ | 4.11 ± 0.03$^b$ | 3.32 ± 0.40$^a$ | 3.48 ± 0.04$^a$ | *0.476* | *0.000* | *0.579* |

Data are given as the mean (n=3) ± SD. In each line, different superscript letters indicate significant differences among treatments ($p \leq 0.05$).
FM30/FO15 = 300g kg$^{-1}$ fishmeal (FM), 150 g kg$^{-1}$ fish oil (FO); FM10/FO3 = 100g kg$^{-1}$ FM; 30g kg$^{-1}$ FO. LD, low rearing density; HD, high rearing density.
PER = Protein efficiency ratio = ((FBW-IBW)/protein intake).
GPE = Gross protein efficiency = 100*[(%final body protein*FBW) - (%initial body protein*IBW)]/total protein intake fish.
GLE = Gross lipid efficiency = 100*[(%final body lipid*FBW) - (%initial body lipid*IBW)]/total lipid intake fish.
LER = Lipid efficiency ratio = ((FBW-IBW)/lipid intake).

**Table 4.** Specific (U mg protein$^{-1}$) digestive enzyme activities of pancreatic (stomach and anterior intestine, AI) and intestinal brush border enzymes of gilthead sea bream reared at low (LD) and high (HD) stocking density and fed the experimental diets over 98 days.

| | *Experimental diets* | | | | *P-value* | | |
|---|---|---|---|---|---|---|---|
| | FM30/FO15 | | FM10/FO3 | | | | |
| | LD | HD | LD | HD | *Density* | *Diet* | *Inter.* |
| *Pancreatic (Stomach/AI)* | | | | | | | |
| Pepsin | 0.33 ± 0.11 | 0.34 ± 0.10 | 0.27 ±0.18 | 0.55 ± 0.20 | *0.157* | *0.414* | *0.165* |
| Trypsin | 0.07 ± 0.03 | 0.04 ± 0.02 | 0.02±0.02 | 0.03 ± 0.01 | *0.225* | *0.053* | *0.225* |
| Chymotrypsin | 0.60 ± 0.06 | 0.31 ± 0.17 | 0.34±0.41 | 0.30 ± 0.20 | *0.276* | *0.366* | *0.413* |
| Total alkaline proteases | 0.56 ± 0.15 | 0.33 ± 0.15 | 0.25±0.28 | 0.27 ± 0.13 | *0.333* | *0.119* | *0.270* |
| Alpha-amylase | 4.49 ± 1.47 | 3.38 ± 0.82 | 3.90±3.24 | 2.37 ± 1.32 | *0.271* | *0.496* | *0.856* |
| Bile salt activated lipase | 0.017± 0.01 | 0.017 ± 0.01 | 0.022±0.02 | 0.025 ± 0.01 | *0.784* | *0.264* | *0.819* |
| *Brush border AI* | | | | | | | |
| Aminopeptidase-N | 0.021±0.01 | 0.022 ± 0.02 | 0.012 ± 0.01 | 0.008 ± 0.01 | *0.816* | *0.128* | *0.722* |
| Phosphatase alkaline | 1.83±0.91 | 1.69 ± 0.31 | 1.10 ± 0.43 | 0.97 ± 0.09 | *0.701* | *0.075* | *0.981* |
| Maltase | 126.4±25.8 | 124.1 ± 35.9 | 122.6 ± 36.9 | 64.9 ± 8.0 | *0.157* | *0.140* | *0.186* |
| LAP | 33.0±3.1$^{ab}$ | 62.3 ± 18.7$^{b}$ | 24.7 ± 6.8$^{a}$ | 41.3 ± 4.8$^{ab}$ | *0.011* | *0.065* | *0.374* |
| *Brush Border PI* | | | | | | | |
| Aminopeptidase | 0.043 ± 0.01$^{b}$ | 0.026 ±0.005$^{ab}$ | 0.0260±0.005$^{ab}$ | 0.021±0.005$^{a}$ | *0.031* | *0.031* | *0.169* |
| Phosphatase alkaline | 0.49 ± 0.10 | 0.94 ± 1.13 | 0.22 ± 0.08 | 0.13 ± 0.02 | *0.600* | *0.137* | *0.432* |
| Maltase | 130.5 ± 70.1 | 164.7 ± 62.9 | 64.8 ± 13.2 | 73.2 ± 26.1 | *0.524* | *0.042* | *0.700* |
| LAP | 46.6 ± 8.1$^{ab}$ | 45.9 ± 1.9$^{ab}$ | 55.6 ±5.9$^{b}$ | 41.8 ± 0.9$^{a}$ | *0.038* | *0.430* | *0.058* |

Data are given as the mean (n = 3) ± SD. In each line, different superscript letters indicate significant differences among treatments ($p \leq 0.05$).
FM30/FO15 = 300g kg$^{-1}$ fishmeal (FM), 150 g kg$^{-1}$ fish oil (FO); FM10/FO3 = 100g kg$^{-1}$ FM; 30g kg$^{-1}$ FO. LD, low rearing density; HD, high rearing density, AI, anterior intestine; PI posterior intestine; LAP, leucine-alanine peptidase.

**Table 5.** Plasma biochemistry values for sea bream kept under high (HD) and low (LD) rearing density and fed the experimental diets.

| | *Experimental diets* | | | | *P - value* | | |
|---|---|---|---|---|---|---|---|
| | FM30/FO15 | | FM10/FO3 | | | | |
| Parameters | LD | HD | LD | HD | Density | Diet | Interaction |
| Glucose (mg dL$^{-1}$) | 119±26 | 123±29 | 117±31 | 101±24 | 0.374 | 0.079 | 0.145 |
| Urea (mg dL$^{-1}$) | 10.7±2.0$^{ab}$ | 9.25±1.44$^{a}$ | 11.6±2.1$^{bc}$ | 13.5±2.8$^{c}$ | 0.760 | 0.000 | 0.003 |
| Creatine (mg dL$^{-1}$) | 0.37±0.14$^{b}$ | 0.30±0.10$^{b}$ | 0.22±0.04$^{a}$ | 0.21±0.04$^{a}$ | 0.169 | 0.000 | 0.090 |
| Uric acid (mg dL$^{-1}$) | 0.51±0.40 | 0.39±0.25 | 0.42±0.42 | 0.32±0.30 | 0.206 | 0.361 | 0.868 |
| Tot bil (mg dL$^{-1}$) | 0.02±0.02 | 0.03±0.01 | 0.04±0.03 | 0.07±0.13 | 0.368 | 0.063 | 0.606 |
| Bil. Ac. (µmol dL$^{-1}$) | 69.3±39.7 | 64.8±41.7 | 48.9±30.4 | 61.2±40.8 | 0.685 | 0.215 | 0.381 |
| Amylase (U L$^{-1}$) | 2.88±5.35 | 0.88±0.34 | 1.25±1.00 | 1.50±2.12 | 0.226 | 0.488 | 0.121 |
| Lipase (U L$^{-1}$) | 2.20±2.43$^{a}$ | 1.69±1.74$^{a}$ | 4.13±2.92$^{ab}$ | 5.22±3.62$^{b}$ | 0.602 | 0.000 | 0.289 |
| CHOL (mg dL$^{-1}$) | 311±75$^{b}$ | 287±71$^{b}$ | 195±27$^{a}$ | 171±35$^{a}$ | 0.089 | 0.000 | 0.987 |
| TRIG (mg dL$^{-1}$) | 792±276 | 793±374 | 810±241 | 830±327 | 0.892 | 0.720 | 0.903 |
| TP (mg dL$^{-1}$) | 4.26±0.76$^{b}$ | 4.10±0.71$^{ab}$ | 3.78±0.29$^{ab}$ | 3.59±0.41$^{a}$ | 0.213 | 0.001 | 0.909 |
| ALB (g dL$^{-1}$) | 0.97±0.19$^{b}$ | 0.90±0.15$^{ab}$ | 0.89±0.06$^{ab}$ | 0.84±0.10$^{a}$ | 0.081 | 0.040 | 0.724 |
| AST (U L$^{-1}$) | 49.2±31.1 | 43.0±32.4 | 55.5±40.8 | 53.3±26.3 | 0.606 | 0.310 | 0.808 |
| ALT (U L$^{-1}$) | 1.81±1.76 | 1.31±0.60 | 1.19±0.54 | 1.11±0.32 | 0.232 | 0.088 | 0.378 |
| ALP (U L$^{-1}$) | 493±190 | 555±265 | 597±259 | 594±274 | 0.632 | 0.251 | 0.601 |
| CK (U L$^{-1}$) | 226±295 | 118±66 | 112±91 | 117±89 | 0.204 | 0.155 | 0.159 |
| GGT (U L$^{-1}$) | 0.10±0.00 | 0.10±0.00 | 0.10±0.00 | 0.10±0.0 | 1.000 | 1.000 | 1.000 |
| LDH (U L$^{-1}$) | 519±662 | 406±409 | 530±646 | 719±527 | 0.792 | 0.259 | 0.292 |
| Ca$^{+2}$ (mg dL$^{-1}$) | 15.0±1.7$^{b}$ | 14.7±1.2$^{ab}$ | 14.3±0.7$^{ab}$ | 13.8±0.9$^{a}$ | 0.142 | 0.008 | 0.670 |
| P (mg dL$^{-1}$) | 13.3±2.1 | 12.0±1.8 | 12.2±1.4 | 12.3±2.4 | 0.249 | 0.381 | 0.183 |
| K$^{+}$ (mEq L$^{-1}$) | 7.16±2.45$^{b}$ | 5.28±1.58$^{a}$ | 7.06±1.70$^{ab}$ | 8.33±2.0$^{b}$ | 0.530 | 0.003 | 0.002 |
| Na$^{+}$ (mEq L$^{-1}$) | 188±6$^{a}$ | 189±5$^{ab}$ | 194±6$^{b}$ | 191±5$^{ab}$ | 0.566 | 0.005 | 0.094 |
| Fe (µg dL$^{-1}$) | 135±33 | 111±28 | 124±30 | 127±37 | 0.206 | 0.766 | 0.090 |
| Cl (mEq L$^{-1}$) | 148±4$^{a}$ | 150±4$^{a}$ | 157±5$^{b}$ | 156±4$^{b}$ | 0.325 | 0.000 | 0.131 |
| Mg (mg dL$^{-1}$) | 4.97±0.98$^{b}$ | 4.30±0.78$^{ab}$ | 3.86±0.50$^{a}$ | 3.86±0.72$^{a}$ | 0.078 | 0.000 | 0.073 |
| UIBC (µg dL$^{-1}$) | 464±78 | 433±97 | 502±68 | 488±96 | 0.300 | 0.031 | 0.695 |
| TIBC (µg dL$^{-1}$) | 599±97 | 544±116 | 626±74 | 616±105 | 0.193 | 0.049 | 0.373 |
| Cortisol (µg dL$^{-1}$) | 3.11±1.74 | 3.78±2.87 | 4.45±3.26 | 4.25±3.99 | 0.837 | 0.244 | 0.278 |
| ALB/GLOB | 0.30±0.03$^{ab}$ | 0.28±0.02$^{a}$ | 0.31±0.02$^{b}$ | 0.31±0.02$^{b}$ | 0.174 | 0.002 | 0.158 |
| CaxP | 201±50 | 178±39 | 175±24 | 169±36 | 0.138 | 0.068 | 0.366 |
| Na/K | 28.9±8.8$^{a}$ | 38.8±10.7$^{b}$ | 29.1±7.5$^{a}$ | 24.1±6.0$^{a}$ | 0.243 | 0.001 | 0.001 |

Data are given as the mean (n=15) ± SD. Different letters indicate significant difference ($P \leq 0.05$) between treatments. FM30/FO15 = 300g kg$^{-1}$ fishmeal (FM), 150 g kg$^{-1}$ fish oil (FO); FM10/FO3 = 100g kg$^{-1}$ FM; 30g kg$^{-1}$ FO. LD, low rearing density; HD, high rearing density. Tot Bil, total bilirubin; CHOL, cholesterol; TRIG, triglycerides; TP, total protein; ALB, albumin; AST, aspartate aminotransferase; ALT, alanine transaminase; ALP, alkaline phosphatase; GGT, gamma-glutamyl transferase; CK, creatine kinase; LDH, lactate dehydrogenase, Ca$^{+2}$, calcium; P, inorganic phosphorus; K$^{+}$, potassium; Na$^{+}$, sodium; Fe, iron; Cl, chloride; Mg, magnesium; UIBC, unsaturated iron binding capacity; TIBC, total iron binding capacity; GLOB, globuline.

**Key to Figures**

Figure 1. A, Lysozyme (U mL$^{-1}$); B, protease activity (%); C, antiprotease activity (%); D, total protein (mg mL$^{-1}$) in skin mucus of gilthead seabream reared at low (LD, light grey) and high (HD, dark grey) stocking density and fed the experimental diets over 98 days. FM30/FO15 = 300g kg$^{-1}$ fishmeal (FM), 150 g kg$^{-1}$ fish oil (FO); FM10/FO3 = 100g kg$^{-1}$ FM; 30g kg$^{-1}$ FO. Data represent the mean ± S.D. (N=24). Different letters denote significant differences between experimental groups ($p < 0.05$).

Figure 2. Barplots representing the sea bream gut bacterial community at two phylogenetic levels: A) phylum; B) Family. In panel C) are reported the boxplots with the families showing a significant difference in relative abundance among groups (p value < 0.05, Wilcoxon ran-sum test; FDR correction). FM30/FO15 = 300g kg$^{-1}$ fishmeal (FM), 150 g kg$^{-1}$ fish oil (FO); FM10/FO3 = 100g kg$^{-1}$ FM; 30g kg$^{-1}$ FO. LD, low rearing density; HD, high rearing density.

Figure 3. Internal biodiversity of sea bream gut microbiota in both feeding regimen and rearing densities computed using Hill numbers (A) highlighted a significant difference between diets ($p < 0.05$; Wilconxon ran-sum test). Principal Coordinates Analysis (PCoA) plots obtained using weighted (B) and unweighted UniFrac (C) showing a significant difference among groups ($p < 0.01$; except FM30/FO15$_{HD}$ *vs* FM30/FO15$_{LD}$, $p > 0.05$; permutation test with pseudo-F ratios, Adonis). FM30/FO15 = 300g kg$^{-1}$ fishmeal (FM), 150 g kg$^{-1}$ fish oil (FO); FM10/FO3 = 100g kg$^{-1}$ FM; 30g kg$^{-1}$ FO. LD, low rearing density; HD, high rearing density.

Figure 1

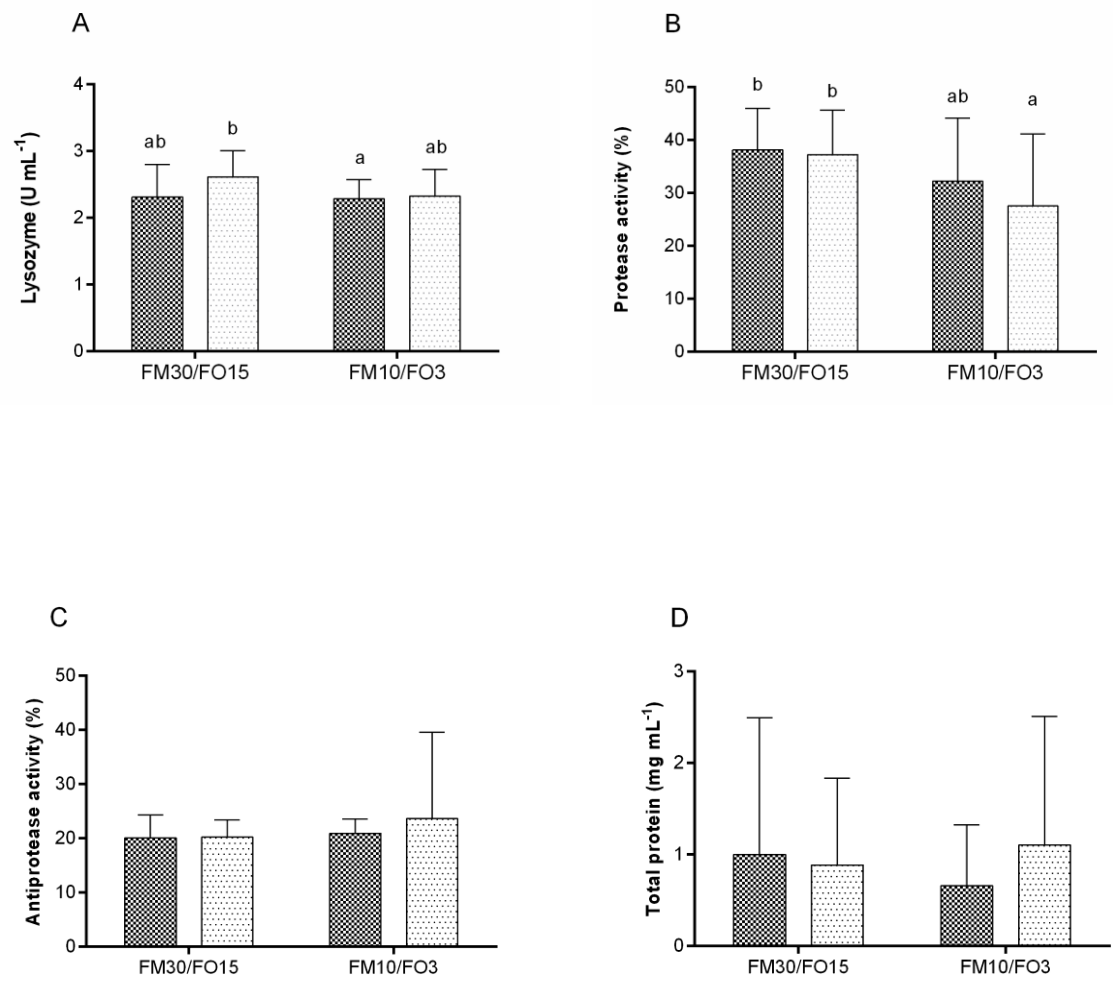

Figure 2

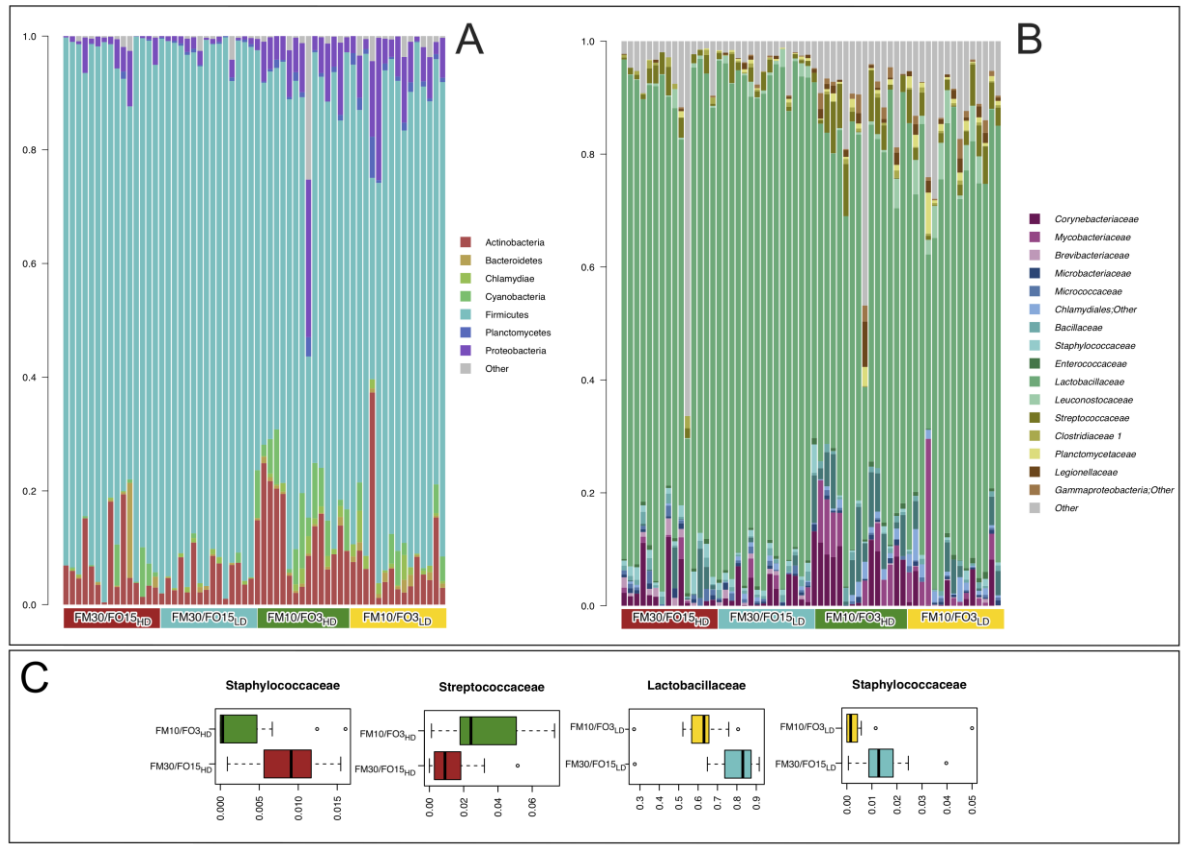

Figure 3

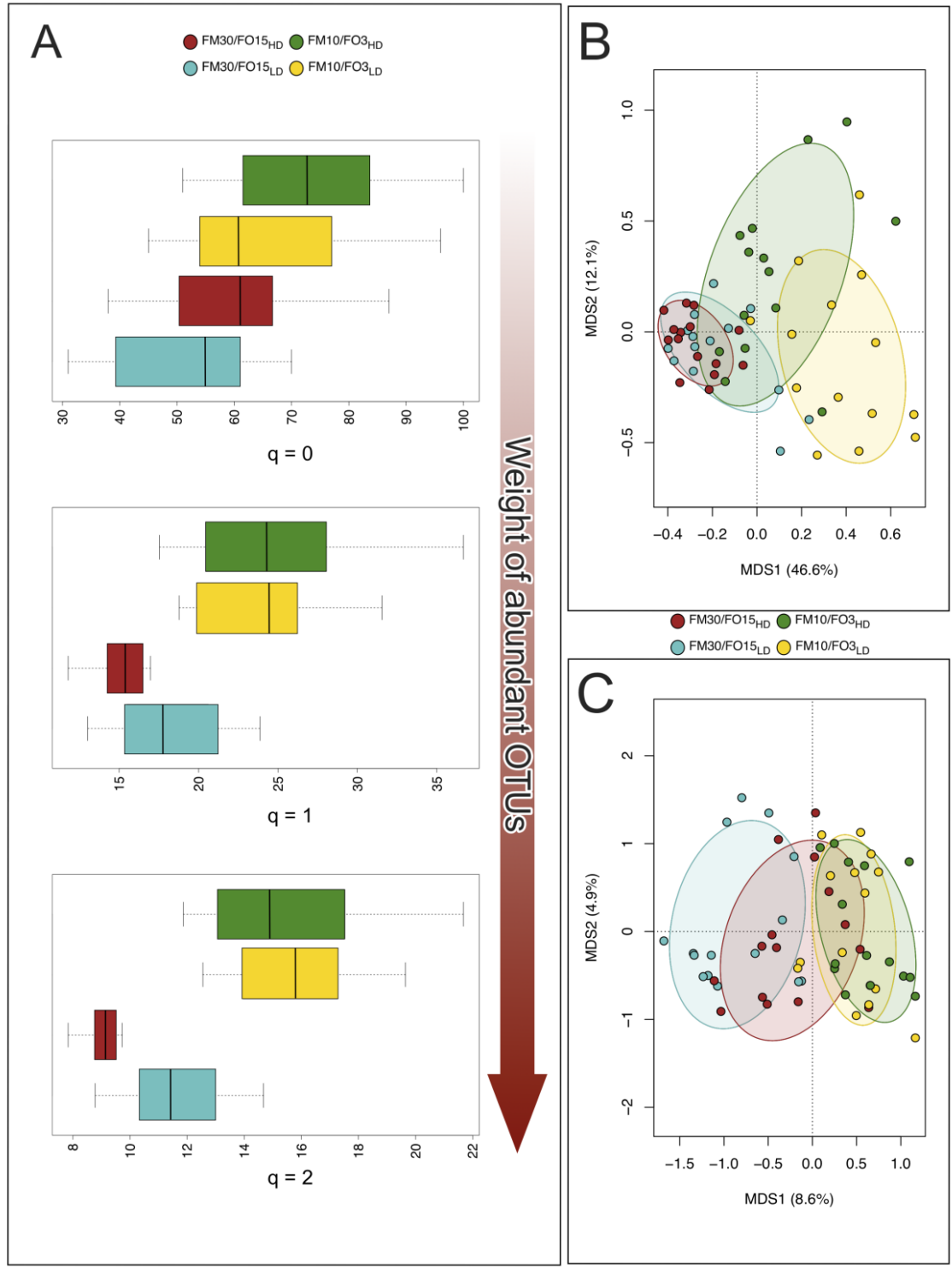

**Supplementary Table 1.**
Mean relative abundance (%) ± SD (n=15) of bacterial phyla, classes, orders, families and genera detected in the distal intestine content of gilthead sea bream fed different diets under high and low rearing density. FM30/FO15 = 300g kg$^{-1}$ fishmeal (FM), 150 g kg$^{-1}$ fish oil (FO); FM10/FO3 = 100g kg$^{-1}$ FM; 30g kg$^{-1}$ FO. LD, low rearing density; HD, high rearing density. Only taxa with mean relative abundance ≥ 0.1% in at least 1 treatment were included.

| Diet | FM30/FO15$_{HD}$ | | FM30/FO15$_{LD}$ | | FM10/FO3$_{HD}$ | | FM10/FO3$_{LD}$ | |
|---|---|---|---|---|---|---|---|---|
| Phylum | Mean | SD | Mean | SD | Mean | SD | Mean | SD |
| *Actinobacteria* | **6.7** | 6.0 | **5.0** | 3.0 | **12.5** | 7.1 | **7.8** | 8.9 |
| *Bacteroidetes* | **1.4** | 4.3 | **0.2** | 0.3 | **0.5** | 0.4 | **0.9** | 0.7 |
| *Chlamydiae* | **0.1** | 0.2 | **0.4** | 0.4 | **1.1** | 1.0 | **1.6** | 1.6 |
| *Chloroflexi* | **0.0** | 0.0 | **0.0** | 0.1 | **0.1** | 0.2 | **0.1** | 0.2 |
| *Cyanobacteria* | **1.5** | 2.8 | **0.2** | 0.5 | **5.4** | 4.0 | **1.9** | 2.5 |
| *Firmicutes* | **87.3** | 9.4 | **92.2** | 4.3 | **69.9** | 13.4 | **77.9** | 13.7 |
| *Gracilibacteria* | **0.0** | 0.0 | **0.1** | 0.3 | **0.2** | 0.9 | **0.0** | 0.0 |
| *Lentisphaerae* | **0.1** | 0.2 | **0.0** | 0.0 | **0.0** | 0.0 | **0.0** | 0.0 |
| *Planctomycetes* | **0.1** | 0.3 | **0.3** | 0.2 | **0.8** | 0.9 | **1.4** | 1.7 |
| *Proteobacteria* | **2.5** | 2.9 | **1.2** | 0.9 | **7.6** | 6.3 | **7.1** | 6.1 |
| *Saccharibacteria* | **0.3** | 0.5 | **0.2** | 0.4 | **0.6** | 1.3 | **0.8** | 1.1 |
| *Spirochaetae* | **0.1** | 0.4 | **0.0** | 0.0 | **0.5** | 1.7 | **0.0** | 0.0 |
| *TM6 (Dependentiae)* | **0.0** | 0.0 | **0.0** | 0.0 | **0.1** | 0.3 | **0.1** | 0.2 |
| *Verrucomicrobia* | **0.0** | 0.1 | **0.0** | 0.2 | **0.1** | 0.3 | **0.0** | 0.0 |
| WS6 | **0.0** | 0.0 | **0.0** | 0.1 | **0.2** | 0.7 | **0.0** | 0.1 |
| Unassigned;Other | **0.0** | 0.1 | **0.1** | 0.2 | **0.2** | 0.8 | **0.2** | 0.4 |
| *Class* | | | | | | | | |
| *Acidimicrobiia* | **0.0** | 0.0 | **0.1** | 0.1 | **0.5** | 0.9 | **0.9** | 1.6 |
| *Actinobacteria* | **6.0** | 5.6 | **4.9** | 3.0 | **11.6** | 7.1 | **6.7** | 7.7 |
| *Coriobacteriia* | **0.4** | 0.9 | **0.0** | 0.1 | **0.4** | 0.8 | **0.1** | 0.3 |
| *Thermoleophilia* | **0.3** | 1.0 | **0.0** | 0.0 | **0.1** | 0.2 | **0.0** | 0.0 |
| *Bacteroidia* | **1.3** | 4.3 | **0.1** | 0.2 | **0.1** | 0.1 | **0.4** | 0.5 |
| *Flavobacteriia* | **0.1** | 0.2 | **0.1** | 0.1 | **0.3** | 0.2 | **0.4** | 0.5 |
| *Sphingobacteriia* | **0.0** | 0.0 | **0.0** | 0.0 | **0.2** | 0.3 | **0.2** | 0.2 |
| *Chlamydiae* | **0.1** | 0.2 | **0.4** | 0.4 | **1.1** | 1.0 | **1.6** | 1.6 |
| *Chloroflexi;KD4-96* | **0.0** | 0.0 | **0.0** | 0.1 | **0.1** | 0.2 | **0.1** | 0.2 |
| *Chloroplast* | **1.5** | 2.8 | **0.2** | 0.5 | **5.4** | 4.0 | **1.9** | 2.5 |
| *Bacilli* | **83.6** | 16.2 | **91.1** | 4.4 | **68.2** | 13.1 | **75.5** | 13.4 |
| *Clostridia* | **3.2** | 7.0 | **1.0** | 0.5 | **1.5** | 1.6 | **2.1** | 1.2 |
| *Erysipelotrichia* | **0.2** | 0.6 | **0.0** | 0.0 | **0.0** | 0.0 | **0.1** | 0.1 |
| *Negativicutes* | **0.3** | 0.8 | **0.0** | 0.1 | **0.2** | 0.3 | **0.3** | 0.4 |
| Gracilibacteria;Other | **0.0** | 0.0 | **0.1** | 0.3 | **0.2** | 0.9 | **0.0** | 0.0 |
| *Planctomycetacia* | **0.1** | 0.3 | **0.3** | 0.2 | **0.8** | 0.9 | **1.4** | 1.7 |
| *Alphaproteobacteria* | **0.6** | 1.1 | **0.3** | 0.3 | **2.5** | 2.3 | **2.0** | 2.6 |
| *Betaproteobacteria* | **0.2** | 0.4 | **0.1** | 0.1 | **0.5** | 0.7 | **0.1** | 0.2 |
| *Deltaproteobacteria* | **0.1** | 0.4 | **0.0** | 0.0 | **0.0** | 0.0 | **0.0** | 0.0 |
| *Epsilonproteobacteria* | **0.1** | 0.2 | **0.0** | 0.0 | **0.0** | 0.0 | **0.2** | 0.6 |
| *Gammaproteobacteria* | **1.5** | 2.4 | **0.8** | 0.6 | **4.5** | 4.5 | **4.8** | 5.8 |
| *Saccharibacteria;uncultured bacterium* | **0.3** | 0.5 | **0.2** | 0.4 | **0.6** | 1.3 | **0.8** | 1.1 |
| *Spirochaetes* | **0.1** | 0.4 | **0.0** | 0.0 | **0.5** | 1.7 | **0.0** | 0.0 |
| *TM6 (Dependentiae);uncultured bacterium* | **0.0** | 0.0 | **0.0** | 0.0 | **0.1** | 0.3 | **0.1** | 0.2 |
| *Verrucomicrobiae* | **0.0** | 0.0 | **0.0** | 0.2 | **0.1** | 0.3 | **0.0** | 0.0 |
| ***Order*** | | | | | | | | |
| *Acidimicrobiales* | **0.0** | 0.0 | **0.1** | 0.1 | **0.5** | 0.9 | **0.9** | 1.6 |
| *Bifidobacteriales* | **0.2** | 0.5 | **0.1** | 0.1 | **0.3** | 0.7 | **0.3** | 0.4 |
| *Corynebacteriales* | **3.4** | 4.8 | **3.0** | 3.0 | **10.0** | 7.1 | **5.2** | 7.8 |
| *Micrococcales* | **2.3** | 1.4 | **1.5** | 0.6 | **0.8** | 0.6 | **0.9** | 0.4 |
| *Propionibacteriales* | **0.0** | 0.1 | **0.2** | 0.2 | **0.4** | 1.1 | **0.3** | 0.3 |
| *Streptomycetales* | **0.0** | 0.0 | **0.0** | 0.1 | **0.0** | 0.0 | **0.0** | 0.0 |
| *Coriobacteriales* | **0.4** | 0.9 | **0.0** | 0.1 | **0.4** | 0.8 | **0.1** | 0.3 |
| *Solirubrobacterales* | **0.2** | 0.8 | **0.0** | 0.0 | **0.1** | 0.2 | **0.0** | 0.0 |
| *Bacteroidales* | **1.3** | 4.3 | **0.1** | 0.2 | **0.1** | 0.1 | **0.4** | 0.5 |
| *Flavobacteriales* | **0.1** | 0.2 | **0.1** | 0.1 | **0.3** | 0.2 | **0.4** | 0.5 |
| *Sphingobacteriales* | **0.0** | 0.0 | **0.0** | 0.0 | **0.2** | 0.3 | **0.2** | 0.2 |
| *Chlamydiales* | **0.1** | 0.2 | **0.4** | 0.4 | **1.1** | 1.0 | **1.6** | 1.6 |
| *Chloroflexi;KD4-96;uncultured bacterium* | **0.0** | 0.0 | **0.0** | 0.1 | **0.1** | 0.2 | **0.1** | 0.2 |
| *Chloroplast;Other* | **1.5** | 2.8 | **0.2** | 0.5 | **5.4** | 4.0 | **1.9** | 2.5 |
| *Bacillales* | **2.8** | 1.1 | **1.8** | 0.7 | **1.7** | 1.2 | **1.1** | 0.7 |
| *Lactobacillales* | **80.8** | 15.9 | **89.2** | 4.2 | **66.4** | 12.7 | **74.4** | 13.1 |
| *Clostridiales* | **3.2** | 7.0 | **1.0** | 0.5 | **1.5** | 1.6 | **2.1** | 1.2 |
| *Erysipelotrichales* | **0.2** | 0.6 | **0.0** | 0.0 | **0.0** | 0.0 | **0.1** | 0.1 |
| *Selenomonadales* | **0.3** | 0.8 | **0.0** | 0.1 | **0.2** | 0.3 | **0.3** | 0.4 |
| *Gracilibacteria;Other* | **0.0** | 0.0 | **0.1** | 0.3 | **0.2** | 0.9 | **0.0** | 0.0 |
| *Planctomycetales* | **0.1** | 0.3 | **0.3** | 0.2 | **0.8** | 0.9 | **1.4** | 1.7 |
| *Rhizobiales* | **0.2** | 0.8 | **0.2** | 0.2 | **1.5** | 2.0 | **1.2** | 2.5 |
| *Rhodobacterales* | **0.0** | 0.1 | **0.0** | 0.1 | **0.2** | 0.3 | **0.3** | 0.4 |
| *Rhodospirillales* | **0.0** | 0.0 | **0.0** | 0.0 | **0.1** | 0.1 | **0.3** | 0.9 |
| *Rickettsiales* | **0.3** | 0.6 | **0.1** | 0.2 | **0.6** | 0.8 | **0.2** | 0.4 |
| *Sphingomonadales* | **0.0** | 0.2 | **0.0** | 0.0 | **0.1** | 0.2 | **0.1** | 0.1 |
| *Burkholderiales* | **0.2** | 0.4 | **0.1** | 0.1 | **0.5** | 0.7 | **0.1** | 0.2 |
| *Campylobacterales* | **0.1** | 0.2 | **0.0** | 0.0 | **0.0** | 0.0 | **0.2** | 0.6 |
| *Aeromonadales* | **0.5** | 1.6 | **0.0** | 0.0 | **0.0** | 0.0 | **0.0** | 0.0 |
| *Enterobacteriales* | **0.3** | 0.7 | **0.1** | 0.1 | **0.2** | 0.3 | **0.2** | 0.3 |
| *Gammaproteobacteria;HTA4* | **0.1** | 0.2 | **0.0** | 0.1 | **0.4** | 1.0 | **0.3** | 0.6 |
| *Legionellales* | **0.2** | 0.2 | **0.3** | 0.4 | **1.9** | 2.7 | **1.4** | 1.1 |
| *Pseudomonadales* | **0.1** | 0.3 | **0.1** | 0.1 | **0.2** | 0.3 | **0.2** | 0.2 |
| *Vibrionales* | **0.3** | 1.2 | **0.0** | 0.1 | **0.3** | 0.6 | **1.6** | 6.1 |
| *Xanthomonadales* | **0.0** | 0.0 | **0.0** | 0.0 | **0.2** | 0.2 | **0.2** | 0.3 |
| *Gammaproteobacteria;Other* | **0.0** | 0.1 | **0.1** | 0.2 | **1.2** | 1.1 | **0.8** | 1.0 |
| *Saccharibacteria;uncultured bacterium* | **0.3** | 0.5 | **0.2** | 0.4 | **0.6** | 1.3 | **0.8** | 1.1 |
| *Spirochaetales* | **0.1** | 0.4 | **0.0** | 0.0 | **0.5** | 1.7 | **0.0** | 0.0 |
| *TM6 (Dependentiae);uncultured bacterium;* | **0.0** | 0.0 | **0.0** | 0.0 | **0.1** | 0.3 | **0.1** | 0.2 |
| *Verrucomicrobiales* | **0.0** | 0.0 | **0.0** | 0.2 | **0.1** | 0.3 | **0.0** | 0.0 |
| *WS6;Other* | **0.0** | 0.0 | **0.0** | 0.1 | **0.1** | 0.6 | **0.0** | 0.0 |
| *Unassigned;Other* | **0.0** | 0.1 | **0.1** | 0.2 | **0.2** | 0.8 | **0.2** | 0.4 |
| *Family* | | | | | | | | |
| *Acidimicrobiales; OM1 clade* | **0.0** | 0.0 | **0.0** | 0.0 | **0.2** | 0.4 | **0.4** | 0.8 |
| *Acidimicrobiales; uncultured* | **0.0** | 0.0 | **0.0** | 0.0 | **0.1** | 0.2 | **0.4** | 1.3 |
| *Bifidobacteriaceae* | **0.2** | 0.5 | **0.1** | 0.1 | **0.3** | 0.7 | **0.3** | 0.4 |
| *Corynebacteriaceae* | **2.9** | 4.1 | **2.4** | 2.6 | **6.1** | 5.2 | **1.8** | 2.9 |
| *Mycobacteriaceae* | **0.4** | 1.0 | **0.6** | 1.2 | **3.9** | 3.9 | **3.3** | 7.5 |
| *Brevibacteriaceae* | **0.8** | 0.9 | **0.4** | 0.5 | **0.0** | 0.1 | **0.1** | 0.2 |
| *Dermabacteraceae* | **0.1** | 0.2 | **0.1** | 0.2 | **0.0** | 0.0 | **0.0** | 0.1 |
| *Intrasporangiaceae* | **0.0** | 0.1 | **0.1** | 0.1 | **0.1** | 0.1 | **0.0** | 0.1 |
| *Microbacteriaceae* | **0.5** | 0.5 | **0.3** | 0.3 | **0.5** | 0.4 | **0.4** | 0.3 |
| *Micrococcaceae* | **0.9** | 0.8 | **0.6** | 0.4 | **0.2** | 0.3 | **0.3** | 0.3 |
| *Nocardioidaceae* | **0.0** | 0.0 | **0.0** | 0.0 | **0.3** | 1.0 | **0.0** | 0.0 |
| *Propionibacteriaceae* | **0.0** | 0.1 | **0.2** | 0.2 | **0.1** | 0.2 | **0.2** | 0.3 |
| *Coriobacteriaceae* | **0.4** | 0.9 | **0.0** | 0.1 | **0.4** | 0.8 | **0.1** | 0.3 |
| *Solirubrobacterales; Elev-16S-1332* | **0.2** | 0.8 | **0.0** | 0.0 | **0.0** | 0.0 | **0.0** | 0.0 |
| *Bacteroidaceae* | **0.1** | 0.4 | **0.0** | 0.1 | **0.0** | 0.1 | **0.2** | 0.4 |
| *Prevotellaceae* | **0.9** | 3.1 | **0.0** | 0.1 | **0.0** | 0.0 | **0.1** | 0.2 |
| *Flavobacteriaceae* | **0.1** | 0.1 | **0.1** | 0.1 | **0.3** | 0.2 | **0.4** | 0.4 |
| *Chitinophagaceae* | **0.0** | 0.0 | **0.0** | 0.0 | **0.2** | 0.3 | **0.1** | 0.2 |
| *Chlamydiales;Other* | **0.1** | 0.2 | **0.4** | 0.4 | **1.1** | 1.0 | **1.6** | 1.6 |
| *Chloroflexi; KD4-96; uncultured bacterium* | **0.0** | 0.0 | **0.0** | 0.1 | **0.1** | 0.2 | **0.1** | 0.2 |
| *Chloroplast;Other* | **1.5** | 2.8 | **0.2** | 0.5 | **5.4** | 4.0 | **1.9** | 2.5 |
| *Bacillaceae* | **1.0** | 0.7 | **0.8** | 0.5 | **0.9** | 0.6 | **0.6** | 0.4 |
| *Paenibacillaceae* | **0.1** | 0.2 | **0.0** | 0.1 | **0.1** | 0.2 | **0.1** | 0.1 |
| *Planococcaceae* | **0.1** | 0.2 | **0.1** | 0.1 | **0.2** | 0.2 | **0.1** | 0.1 |
| *Staphylococcaceae* | **1.4** | 1.0 | **0.9** | 0.4 | **0.6** | 1.3 | **0.3** | 0.5 |
| *Bacillales;Other* | **0.1** | 0.4 | **0.0** | 0.0 | **0.0** | 0.0 | **0.0** | 0.0 |
| *Aerococcaceae* | **0.1** | 0.1 | **0.1** | 0.1 | **0.0** | 0.1 | **0.0** | 0.1 |
| *Carnobacteriaceae* | **0.0** | 0.1 | **0.0** | 0.1 | **0.1** | 0.2 | **0.1** | 0.2 |
| *Enterococcaceae* | **0.4** | 0.3 | **0.3** | 0.2 | **0.4** | 0.4 | **0.4** | 0.4 |
| *Lactobacillaceae* | **77.9** | 16.1 | **86.5** | 4.4 | **61.3** | 12.4 | **67.6** | 12.2 |
| *Leuconostocaceae* | **0.5** | 0.8 | **1.0** | 1.1 | **0.5** | 1.3 | **3.0** | 2.7 |
| *Streptococcaceae* | **2.0** | 1.5 | **1.3** | 1.4 | **4.1** | 3.7 | **3.2** | 2.3 |
| *Clostridiaceae 1* | **0.7** | 0.7 | **0.2** | 0.2 | **0.4** | 0.4 | **0.4** | 0.2 |
| *Clostridiaceae 2* | **0.0** | 0.0 | **0.0** | 0.0 | **0.1** | 0.5 | **0.0** | 0.0 |
| *Clostridiales;Family XI* | **0.3** | 0.3 | **0.3** | 0.2 | **0.2** | 0.2 | **0.3** | 0.3 |
| *Clostridiales; Family XIII* | **0.1** | 0.2 | **0.1** | 0.3 | **0.0** | 0.0 | **0.0** | 0.0 |
| *Lachnospiraceae* | **0.6** | 2.1 | **0.2** | 0.3 | **0.1** | 0.2 | **0.4** | 0.5 |
| *Peptostreptococcaceae* | **0.2** | 0.3 | **0.1** | 0.1 | **0.1** | 0.2 | **0.1** | 0.2 |
| *Ruminococcaceae* | **1.1** | 4.1 | **0.0** | 0.1 | **0.0** | 0.0 | **0.4** | 0.6 |
| *Clostridiales;Other* | **0.1** | 0.3 | **0.1** | 0.2 | **0.6** | 0.9 | **0.5** | 0.8 |
| *Erysipelotrichaceae* | **0.2** | 0.6 | **0.0** | 0.0 | **0.0** | 0.0 | **0.1** | 0.1 |
| *Acidaminococcaceae* | **0.2** | 0.6 | **0.0** | 0.0 | **0.0** | 0.0 | **0.1** | 0.3 |
| *Veillonellaceae* | **0.2** | 0.3 | **0.0** | 0.1 | **0.2** | 0.3 | **0.1** | 0.2 |
| *Gracilibacteria;Other* | **0.0** | 0.0 | **0.1** | 0.3 | **0.2** | 0.9 | **0.0** | 0.0 |
| *Planctomycetaceae* | **0.1** | 0.3 | **0.3** | 0.2 | **0.8** | 0.9 | **1.4** | 1.7 |
| *Bradyrhizobiaceae* | **0.0** | 0.0 | **0.0** | 0.0 | **0.7** | 1.6 | **0.1** | 0.3 |
| *Brucellaceae* | **0.0** | 0.0 | **0.0** | 0.0 | **0.2** | 0.5 | **0.1** | 0.2 |
| *Hyphomicrobiaceae* | **0.1** | 0.2 | **0.0** | 0.0 | **0.0** | 0.1 | **0.1** | 0.5 |
| *Phyllobacteriaceae* | **0.2** | 0.5 | **0.1** | 0.2 | **0.2** | 0.5 | **0.5** | 1.8 |
| *Rhizobiaceae* | **0.0** | 0.0 | **0.0** | 0.0 | **0.0** | 0.1 | **0.1** | 0.3 |
| *Rhizobiales;Other* | **0.0** | 0.0 | **0.0** | 0.1 | **0.2** | 0.3 | **0.2** | 0.2 |
| *Rhodobacteraceae* | **0.0** | 0.1 | **0.0** | 0.1 | **0.2** | 0.3 | **0.3** | 0.4 |
| *Acetobacteraceae* | **0.0** | 0.0 | **0.0** | 0.0 | **0.0** | 0.0 | **0.2** | 0.9 |
| *Mitochondria* | **0.3** | 0.6 | **0.1** | 0.2 | **0.6** | 0.8 | **0.2** | 0.4 |
| *Sphingomonadaceae* | **0.0** | 0.2 | **0.0** | 0.0 | **0.1** | 0.2 | **0.1** | 0.1 |
| *Comamonadaceae* | **0.2** | 0.4 | **0.0** | 0.1 | **0.3** | 0.4 | **0.1** | 0.2 |
| *Oxalobacteraceae* | **0.0** | 0.0 | **0.0** | 0.1 | **0.2** | 0.4 | **0.0** | 0.1 |
| *Helicobacteraceae* | **0.0** | 0.0 | **0.0** | 0.0 | **0.0** | 0.0 | **0.2** | 0.6 |
| *Aeromonadaceae* | **0.2** | 0.5 | **0.0** | 0.0 | **0.0** | 0.0 | **0.0** | 0.0 |
| *Succinivibrionaceae* | **0.3** | 1.3 | **0.0** | 0.0 | **0.0** | 0.0 | **0.0** | 0.0 |
| *Enterobacteriaceae* | **0.3** | 0.7 | **0.1** | 0.1 | **0.2** | 0.3 | **0.2** | 0.3 |
| *Gammaproteobacteria;HTA4;Other* | **0.1** | 0.2 | **0.0** | 0.1 | **0.4** | 1.0 | **0.3** | 0.6 |
| *Coxiellaceae* | **0.1** | 0.2 | **0.1** | 0.1 | **0.5** | 0.9 | **0.4** | 0.4 |
| *Legionellaceae* | **0.1** | 0.2 | **0.3** | 0.3 | **1.4** | 2.0 | **1.0** | 0.9 |
| *Moraxellaceae* | **0.1** | 0.3 | **0.1** | 0.1 | **0.2** | 0.2 | **0.1** | 0.3 |
| *Vibrionaceae* | **0.3** | 1.2 | **0.0** | 0.1 | **0.3** | 0.6 | **1.6** | 6.1 |
| *Xanthomonadaceae* | **0.0** | 0.0 | **0.0** | 0.0 | **0.1** | 0.2 | **0.2** | 0.3 |
| *Gammaproteobacteria;Other* | **0.0** | 0.1 | **0.1** | 0.2 | **1.2** | 1.1 | **0.8** | 1.0 |
| *Saccharibacteria; uncultured bacterium* | **0.3** | 0.5 | **0.2** | 0.4 | **0.6** | 1.3 | **0.8** | 1.1 |
| *Brevinemataceae* | **0.0** | 0.0 | **0.0** | 0.0 | **0.5** | 1.7 | **0.0** | 0.0 |
| *TM6 (Dependentiae); uncultured bacterium* | **0.0** | 0.0 | **0.0** | 0.0 | **0.1** | 0.3 | **0.1** | 0.2 |
| *Verrucomicrobiaceae* | **0.0** | 0.0 | **0.0** | 0.2 | **0.1** | 0.2 | **0.0** | 0.0 |
| *WS6;Other* | **0.0** | 0.0 | **0.0** | 0.1 | **0.1** | 0.6 | **0.0** | 0.0 |
| *Unassigned;Other* | **0.0** | 0.1 | **0.1** | 0.2 | **0.2** | 0.8 | **0.2** | 0.4 |
| *Genus* | | | | | | | | |
| *Acidimicrobiales; OM1 clade; uncultured bacterium* | **0.0** | 0.0 | **0.0** | 0.0 | **0.2** | 0.4 | **0.4** | 0.8 |
| *Acidimicrobiales; uncultured;Other* | **0.0** | 0.0 | **0.0** | 0.0 | **0.0** | 0.1 | **0.4** | 1.3 |
| *Bifidobacterium* | **0.2** | 0.5 | **0.1** | 0.1 | **0.3** | 0.7 | **0.3** | 0.4 |
| *Corynebacterium 1* | **2.8** | 4.1 | **2.3** | 2.5 | **6.1** | 5.2 | **1.8** | 2.9 |
| *Mycobacterium* | **0.4** | 1.0 | **0.6** | 1.2 | **3.9** | 3.9 | **3.3** | 7.5 |
| *Nocardia* | **0.0** | 0.0 | **0.0** | 0.0 | **0.0** | 0.1 | **0.0** | 0.0 |
| *Brevibacterium* | **0.8** | 0.9 | **0.4** | 0.5 | **0.0** | 0.1 | **0.1** | 0.2 |
| *Brachybacterium* | **0.1** | 0.2 | **0.1** | 0.2 | **0.0** | 0.0 | **0.0** | 0.1 |
| *Intrasporangiaceae;Other* | **0.0** | 0.1 | **0.1** | 0.1 | **0.1** | 0.1 | **0.0** | 0.1 |
| *Leucobacter* | **0.3** | 0.4 | **0.1** | 0.2 | **0.2** | 0.4 | **0.1** | 0.2 |
| *Microbacteriaceae;Other* | **0.2** | 0.5 | **0.2** | 0.3 | **0.3** | 0.4 | **0.2** | 0.3 |
| *Arthrobacter* | **0.2** | 0.4 | **0.3** | 0.4 | **0.1** | 0.2 | **0.1** | 0.2 |
| *Glutamicibacter* | **0.2** | 0.6 | **0.1** | 0.2 | **0.0** | 0.1 | **0.0** | 0.1 |
| *Kocuria* | **0.3** | 0.4 | **0.2** | 0.2 | **0.1** | 0.3 | **0.2** | 0.2 |
| *Micrococcaceae;Other* | **0.1** | 0.1 | **0.0** | 0.0 | **0.0** | 0.1 | **0.0** | 0.1 |
| *Nocardioides* | **0.0** | 0.0 | **0.0** | 0.0 | **0.3** | 1.0 | **0.0** | 0.0 |
| *Propionibacterium* | **0.0** | 0.1 | **0.1** | 0.1 | **0.1** | 0.1 | **0.2** | 0.3 |
| *Collinsella* | **0.2** | 0.8 | **0.0** | 0.0 | **0.0** | 0.0 | **0.0** | 0.1 |
| *Enterorhabdus* | **0.0** | 0.1 | **0.0** | 0.0 | **0.3** | 0.8 | **0.0** | 0.0 |
| *Coriobacteriaceae; uncultured* | **0.1** | 0.5 | **0.0** | 0.0 | **0.0** | 0.0 | **0.0** | 0.0 |
| *Solirubrobacterales; Elev-16S-1332 uncultured bacterium* | **0.2** | 0.8 | **0.0** | 0.0 | **0.0** | 0.0 | **0.0** | 0.0 |
| *Bacteroides* | **0.1** | 0.4 | **0.0** | 0.1 | **0.0** | 0.1 | **0.2** | 0.4 |
| *Bacteroidales S24-7 group; uncultured bacterium* | **0.0** | 0.2 | **0.0** | 0.0 | **0.0** | 0.0 | **0.0** | 0.0 |
| *Prevotella 2* | **0.1** | 0.5 | **0.0** | 0.0 | **0.0** | 0.0 | **0.0** | 0.0 |
| *Prevotella 9* | **0.7** | 2.2 | **0.0** | 0.1 | **0.0** | 0.0 | **0.1** | 0.2 |
| *Cloacibacterium* | **0.0** | 0.0 | **0.0** | 0.0 | **0.0** | 0.0 | **0.0** | 0.1 |
| *Flavobacterium* | **0.0** | 0.1 | **0.0** | 0.0 | **0.1** | 0.2 | **0.2** | 0.2 |
| *Flavobacteriaceae;Other* | **0.0** | 0.1 | **0.0** | 0.1 | **0.1** | 0.2 | **0.1** | 0.3 |
| *Sediminibacterium* | **0.0** | 0.0 | **0.0** | 0.0 | **0.1** | 0.2 | **0.1** | 0.2 |
| *Chlamydiales;Other* | **0.1** | 0.2 | **0.4** | 0.4 | **1.1** | 1.0 | **1.6** | 1.6 |
| *Chloroflexi; KD4-96; uncultured bacterium* | **0.0** | 0.0 | **0.0** | 0.1 | **0.1** | 0.2 | **0.1** | 0.2 |
| *Chloroplast;Other* | **1.5** | 2.8 | **0.2** | 0.5 | **5.4** | 4.0 | **1.9** | 2.5 |
| *Bacillus* | **0.5** | 0.4 | **0.5** | 0.4 | **0.8** | 0.6 | **0.5** | 0.4 |
| *Bacillaceae;Other* | **0.4** | 0.3 | **0.3** | 0.3 | **0.0** | 0.1 | **0.1** | 0.1 |
| *Brevibacillus* | **0.0** | 0.1 | **0.0** | 0.0 | **0.0** | 0.0 | **0.0** | 0.1 |
| *Paenibacillus* | **0.0** | 0.0 | **0.0** | 0.0 | **0.1** | 0.2 | **0.0** | 0.1 |
| *Planococcaceae;Other* | **0.1** | 0.1 | **0.1** | 0.1 | **0.2** | 0.2 | **0.0** | 0.1 |
| *Staphylococcus* | **1.3** | 1.0 | **0.8** | 0.4 | **0.5** | 1.3 | **0.3** | 0.5 |
| *Staphylococcaceae;Other* | **0.1** | 0.2 | **0.1** | 0.2 | **0.1** | 0.2 | **0.0** | 0.1 |
| *Bacillales;Other* | **0.1** | 0.4 | **0.0** | 0.0 | **0.0** | 0.0 | **0.0** | 0.0 |
| *Granulicatella* | **0.0** | 0.1 | **0.0** | 0.1 | **0.1** | 0.2 | **0.1** | 0.2 |
| *Enterococcus* | **0.3** | 0.3 | **0.3** | 0.2 | **0.4** | 0.4 | **0.4** | 0.4 |
| *Lactobacillus* | **77.9** | 16.1 | **86.5** | 4.4 | **61.3** | 12.4 | **67.6** | 12.2 |
| *Leuconostoc* | **0.0** | 0.1 | **0.3** | 0.3 | **0.1** | 0.2 | **0.1** | 0.3 |
| *Weissella* | **0.4** | 0.8 | **0.7** | 0.9 | **0.4** | 1.3 | **2.8** | 2.8 |
| *Lactococcus* | **0.3** | 0.4 | **0.1** | 0.2 | **0.5** | 0.4 | **0.3** | 0.3 |
| *Streptococcus* | **1.6** | 1.6 | **1.2** | 1.4 | **3.6** | 3.5 | **2.9** | 2.2 |
| *Clostridium sensu stricto 1* | **0.4** | 0.7 | **0.1** | 0.1 | **0.0** | 0.1 | **0.1** | 0.1 |
| *Clostridiaceae 1;Other* | **0.2** | 0.4 | **0.1** | 0.2 | **0.3** | 0.4 | **0.2** | 0.3 |
| *Alkaliphilus* | **0.0** | 0.0 | **0.0** | 0.0 | **0.1** | 0.5 | **0.0** | 0.0 |
| *Clostridiales; Family XI;uncultured* | **0.1** | 0.2 | **0.0** | 0.1 | **0.0** | 0.2 | **0.0** | 0.1 |
| *Clostridiales; Family XI;Other* | **0.1** | 0.2 | **0.2** | 0.2 | **0.1** | 0.2 | **0.2** | 0.3 |
| *Blautia* | **0.0** | 0.1 | **0.0** | 0.0 | **0.0** | 0.0 | **0.1** | 0.2 |
| *Roseburia* | **0.2** | 0.6 | **0.0** | 0.2 | **0.0** | 0.0 | **0.0** | 0.0 |
| *Peptostreptococcaceae; Other* | **0.2** | 0.3 | **0.0** | 0.1 | **0.1** | 0.2 | **0.0** | 0.1 |
| *Faecalibacterium* | **0.3** | 1.0 | **0.0** | 0.1 | **0.0** | 0.0 | **0.1** | 0.3 |
| *Ruminococcaceae UCG-002* | **0.1** | 0.6 | **0.0** | 0.0 | **0.0** | 0.0 | **0.0** | 0.1 |
| *Ruminococcaceae UCG-005* | **0.2** | 0.7 | **0.0** | 0.0 | **0.0** | 0.0 | **0.0** | 0.0 |
| *Ruminococcus 2* | **0.1** | 0.4 | **0.0** | 0.1 | **0.0** | 0.0 | **0.0** | 0.1 |
| *[Eubacterium] coprostanoligenes group* | **0.1** | 0.4 | **0.0** | 0.0 | **0.0** | 0.0 | **0.0** | 0.0 |
| *Ruminococcaceae; uncultured* | **0.0** | 0.0 | **0.0** | 0.0 | **0.0** | 0.0 | **0.1** | 0.3 |
| *Clostridiales; Other* | **0.1** | 0.3 | **0.1** | 0.2 | **0.6** | 0.9 | **0.5** | 0.8 |
| *Phascolarctobacterium* | **0.2** | 0.6 | **0.0** | 0.0 | **0.0** | 0.0 | **0.0** | 0.1 |
| *Acidaminococcaceae;Other* | **0.0** | 0.0 | **0.0** | 0.0 | **0.0** | 0.0 | **0.1** | 0.2 |
| *Megasphaera* | **0.0** | 0.1 | **0.0** | 0.1 | **0.2** | 0.2 | **0.1** | 0.1 |
| *Gracilibacteria; Othe* | **0.0** | 0.0 | **0.1** | 0.3 | **0.2** | 0.9 | **0.0** | 0.0 |
| *Planctomycetaceae; Pir4 lineage* | **0.0** | 0.0 | **0.1** | 0.1 | **0.2** | 0.2 | **0.2** | 0.0 |
| *Planctomyces* | **0.0** | 0.1 | **0.1** | 0.2 | **0.3** | 0.3 | **0.6** | 0.0 |
| *Planctomycetaceae; uncultured* | **0.1** | 0.3 | **0.1** | 0.2 | **0.2** | 0.4 | **0.4** | 0.1 |
| *Bradyrhizobium* | **0.0** | 0.0 | **0.0** | 0.0 | **0.7** | 1.6 | **0.1** | 0.0 |
| *Ochrobactrum* | **0.0** | 0.0 | **0.0** | 0.0 | **0.2** | 0.5 | **0.1** | 0.0 |
| *Hyphomicrobium* | **0.1** | 0.2 | **0.0** | 0.0 | **0.0** | 0.0 | **0.1** | 0.0 |
| *Mesorhizobium* | **0.0** | 0.0 | **0.0** | 0.0 | **0.0** | 0.0 | **0.1** | 0.1 |
| *Phyllobacteriaceae; Other* | **0.2** | 0.5 | **0.1** | 0.2 | **0.2** | 0.5 | **0.4** | 0.4 |
| *Rhizobiales; Other* | **0.0** | 0.0 | **0.0** | 0.1 | **0.2** | 0.3 | **0.2** | 0.2 |
| *Rhodobacteraceae; Other* | **0.0** | 0.0 | **0.0** | 0.1 | **0.1** | 0.2 | **0.2** | 0.0 |
| *Acetobacteraceae; Other* | **0.0** | 0.0 | **0.0** | 0.0 | **0.0** | 0.0 | **0.2** | 0.1 |
| *Mitochondria;Other* | **0.3** | 0.6 | **0.1** | 0.2 | **0.6** | 0.8 | **0.2** | 0.0 |
| *Delftia* | **0.0** | 0.0 | **0.0** | 0.1 | **0.3** | 0.4 | **0.1** | 0.0 |
| *Comamonadaceae;Other* | **0.1** | 0.4 | **0.0** | 0.0 | **0.0** | 0.0 | **0.0** | 0.0 |
| *Oxalobacteraceae;Other* | **0.0** | 0.0 | **0.0** | 0.1 | **0.2** | 0.4 | **0.0** | 0.0 |
| *Succinivibrio* | **0.3** | 1.3 | **0.0** | 0.0 | **0.0** | 0.0 | **0.0** | 0.6 |
| *Escherichia-Shigella* | **0.2** | 0.7 | **0.0** | 0.0 | **0.0** | 0.1 | **0.1** | 0.0 |
| *Serratia* | **0.1** | 0.2 | **0.0** | 0.0 | **0.2** | 0.2 | **0.1** | 0.0 |
| *Gammaproteobacteria; HTA4;Other* | **0.1** | 0.2 | **0.0** | 0.1 | **0.4** | 1.0 | **0.3** | 0.0 |
| *Aquicella* | **0.0** | 0.1 | **0.0** | 0.1 | **0.2** | 0.4 | **0.2** | 0.2 |
| *Coxiella* | **0.1** | 0.2 | **0.0** | 0.1 | **0.3** | 0.5 | **0.2** | 0.0 |
| *Legionella* | **0.1** | 0.2 | **0.3** | 0.3 | **1.3** | 2.0 | **0.8** | 0.3 |
| *Legionellaceae; Other* | **0.0** | 0.0 | **0.0** | 0.0 | **0.0** | 0.2 | **0.2** | 0.0 |
| *Acinetobacter* | **0.1** | 0.2 | **0.1** | 0.1 | **0.2** | 0.2 | **0.1** | 0.5 |
| *Photobacterium* | **0.2** | 1.0 | **0.0** | 0.0 | **0.2** | 0.6 | **0.4** | 0.0 |
| *Vibrio* | **0.1** | 0.2 | **0.0** | 0.1 | **0.0** | 0.0 | **1.2** | 0.2 |
| *Stenotrophomonas* | **0.0** | 0.0 | **0.0** | 0.0 | **0.1** | 0.2 | **0.1** | 0.1 |
| *Gammaproteobacteria;Other;* | **0.0** | 0.1 | **0.1** | 0.2 | **1.2** | 1.1 | **0.8** | 4.6 |
| *Saccharibacteria; uncultured bacterium;* | **0.3** | 0.5 | **0.2** | 0.4 | **0.6** | 1.3 | **0.8** | 0.1 |
| *Brevinema* | **0.0** | 0.0 | **0.0** | 0.0 | **0.5** | 1.7 | **0.0** | 0.1 |
| *TM6 (Dependentiae); uncultured bacterium* | **0.0** | 0.0 | **0.0** | 0.0 | **0.1** | 0.3 | **0.1** | 0.1 |
| *WS6;Other* | **0.0** | 0.0 | **0.0** | 0.1 | **0.1** | 0.6 | **0.0** | 0.0 |
| *Unassigned;Other* | **0.0** | 0.1 | **0.1** | 0.2 | **0.2** | 0.8 | **0.2** | 0.0 |